\documentclass[journal=apchd5,manuscript=article,layout=onecolumn]{achemso}
\usepackage{chemformula} 
\usepackage[T1]{fontenc} 
\usepackage[normalem]{ulem}
\usepackage{graphicx}
\usepackage[utf8]{inputenc}
\usepackage{mathptmx}
\usepackage{etoolbox}
\usepackage{float}
\usepackage{upgreek}
\usepackage{caption}

\usepackage[sectionbib]{bibunits}
\defaultbibliographystyle{unsrt} 
\defaultbibliography{references}

\newcommand{\onlinecite}[1]{\hspace{-1 ex} \nocite{#1}\citenum{#1}} 

\author{Daniel~A.~Vajner}
\affiliation[TU Berlin]{Institute of Solid State Physics, Technical University of Berlin, 10623 Berlin, Germany}

\author{Pawe\l{}~Holewa}
\affiliation[Wroclaw]{Department of Experimental Physics, Faculty of Fundamental Problems of Technology, Wroc\l{}aw University of Science and Technology, Wyb. Wyspia\'{n}skiego 27, 50-370 Wroc\l{}aw, Poland}
\alsoaffiliation[DTU]{DTU Electro, Department of Electrical and Photonics Engineering, Technical University of Denmark, Kongens Lyngby 2800, Denmark}
\alsoaffiliation{NanoPhoton - Center for Nanophotonics, Technical University of Denmark, 2800 Kongens Lyngby, Denmark}

\author{Emilia~Zi\k{e}ba-Ost\'{o}j}
\affiliation{Department of Experimental Physics, Faculty of Fundamental Problems of Technology, Wroc\l{}aw University of Science and Technology, Wyb. Wyspia\'{n}skiego 27, 50-370 Wroc\l{}aw, Poland}

\author{Maja~Wasiluk}
\affiliation{Department of Experimental Physics, Faculty of Fundamental Problems of Technology, Wroc\l{}aw University of Science and Technology, Wyb. Wyspia\'{n}skiego 27, 50-370 Wroc\l{}aw, Poland}

\author{Martin~von~Helversen}
\affiliation{Institute of Solid State Physics, Technical University of Berlin, 10623 Berlin, Germany}

\author{Aurimas~Sakanas}
\affiliation{DTU Electro, Department of Electrical and Photonics Engineering, Technical University of Denmark, Kongens Lyngby 2800, Denmark}

\author{Alexander~Huck}
\affiliation{Center for Macroscopic Quantum States (bigQ), Department of Physics, Technical University of Denmark, 2800 Kgs. Lyngby, Denmark}

\author{Kresten~Yvind}
\affiliation{DTU Electro, Department of Electrical and Photonics Engineering, Technical University of Denmark, Kongens Lyngby 2800, Denmark}
\alsoaffiliation{NanoPhoton - Center for Nanophotonics, Technical University of Denmark, 2800 Kongens Lyngby, Denmark}

\author{Niels~Gregersen}
\affiliation{DTU Electro, Department of Electrical and Photonics Engineering, Technical University of Denmark, Kongens Lyngby 2800, Denmark}

\author{Anna~Musia\l{}}
\affiliation{Department of Experimental Physics, Faculty of Fundamental Problems of Technology, Wroc\l{}aw University of Science and Technology, Wyb. Wyspia\'{n}skiego 27, 50-370 Wroc\l{}aw, Poland}

\author{Marcin~Syperek}
\affiliation{Department of Experimental Physics, Faculty of Fundamental Problems of Technology, Wroc\l{}aw University of Science and Technology, Wyb. Wyspia\'{n}skiego 27, 50-370 Wroc\l{}aw, Poland}

\author{Elizaveta~Semenova}
\affiliation{DTU Electro, Department of Electrical and Photonics Engineering, Technical University of Denmark, Kongens Lyngby 2800, Denmark}
\alsoaffiliation{NanoPhoton - Center for Nanophotonics, Technical University of Denmark, 2800 Kongens Lyngby, Denmark}

\author{Tobias~Heindel}
\email{tobias.heindel@tu-berlin.de}
\affiliation{Institute of Solid State Physics, Technical University of Berlin, 10623 Berlin, Germany}

\title{On-Demand Generation of Indistinguishable Photons in the Telecom C-Band Using Quantum Dot Devices}

\begin{document}
\begin{bibunit}

\newpage
\begin{abstract}
 Semiconductor quantum dots (QDs) enable the generation of single and entangled photons, useful for various applications in photonic quantum technologies. Specifically for quantum communication via fiber-optical networks, operation in the telecom C-band centered around 1550$\,$nm is ideal. The direct generation of QD-photons in this spectral range and with high quantum-optical quality, however, remained challenging. Here, we demonstrate the coherent on-demand generation of indistinguishable photons in the telecom C-band from single QD devices consisting of InAs/InP QD-mesa structures heterogeneously integrated with a metallic reflector on a silicon wafer. Using pulsed two-photon resonant excitation of the biexciton-exciton radiative cascade, we observe Rabi rotations up to pulse areas of $4\pi$ and a high single-photon purity in terms of $g^{(2)}(0)=0.005(1)$ and $0.015(1)$ for exciton and biexciton photons, respectively. Applying two independent experimental methods, based on fitting Rabi rotations in the emission intensity and performing photon cross-correlation measurements, we consistently obtain preparation fidelities at the $\pi$-pulse exceeding 80$\%$. Finally, performing Hong-Ou-Mandel-type two-photon interference experiments we obtain a photon-indistinguishability of the full photon wave packet of up to $35(3)\%$, representing a significant advancement in the photon-indistinguishability of single photons emitted directly in the telecom C-band. 
\end{abstract}

\section*{Introduction} 
Single indistinguishable photons are a key resource for many applications in quantum information ranging from quantum communication to distributed quantum computing. They are essential for quantum networks and the quantum internet \cite{kimble2008quantum}. While a plethora of quantum emitters enable the generation of single photons \cite{aharonovich2016solid,chakraborty2019advances}, epitaxial semiconductor quantum dots (QDs) turned out to be advantageous in many regards\cite{senellart2017high,arakawa2020progress,rodt2020deterministically,lu2021quantum}. Over the last decades, photons generated on-demand using QDs demonstrated unprecedented quantum optical properties in terms of high purity, brightness and indistinguishability and have been repeatedly employed in implementations of quantum communication \cite{vajner2022quantum}. So far, such close-to-ideal single-photon sources have only been demonstrated at wavelengths around $780\,\text{nm}$ for GaAs/AlGaAs QDs \cite{schweickert2018demand,scholl2019resonance} and $930\,\text{nm}$ for InGaAs/GaAs QDs \cite{wang2016near,hanschke2018quantum,tomm2021bright}. For long-distance quantum information transfer via optical fibers, however, wavelengths around 1550$\,$nm, i.\,e. in the telecom C-band, are required to benefit from the lowest losses in optical fibers. To shift the emission of QDs to C-band wavelengths, quantum frequency conversion of QD-photons emitted at shorter wavelengths can be used \cite{kambs2016low,da2022pure}, which however introduces additional conversion losses ultimately limiting the source efficiency. 
For this reason, QDs directly emitting photons at telecom wavelengths are desirable, requiring carefully tailored growth schemes.

One solution is to introduce metamorphic buffer layers to engineer the strain and size of InAs/InGaAs QDs, which shifts the emission to longer wavelengths \cite{semenova2008metamorphic}. QDs for the telecom C-band grown on metamorphic buffers have been advanced in recent years \cite{portalupi2019inas,zeuner2021demand} and recently triggered photon-indistinguishabilities of $V \approx 10 \%$ and $V = 14(2)\%$ were reported under quasi-resonant \cite{nawrath2023bright} and resonant \cite{nawrath2021resonance} excitation, respectively. An alternative approach uses the InP material system to grow InAs/InP QDs naturally emitting at telecom C-band wavelengths. Here, no additional metamorphic buffer layer is required, which enables high crystalline quality of the QD material and reduces the complexity of the growth in favor of improved device scalability. Various studies investigated the properties of InP-based QDs under above-barrier and quasi-resonant excitation \cite{birowosuto2012fast,miyazawa2016single,muller2018quantum,musial2021inp,anderson2021coherence,holewa2022bright}. A notable recent advancement in this context concerns the demonstration of a scalable device platform, by deterministically integrating single InAs/InP QDs into circular Bragg grating cavities, resulting in a triggered, Purcell-enhanced emission and a photon-indistinguishability of $V=19(3)\%$ under quasi-resonant excitation \cite{holewa2023scalable}. The pulsed coherent generation of indistinguishable single photons from QDs at C-band wavelengths, however, is an important requirement for applications that had not been achieved to date. Pulsed photon generation is indispensable for on-demand operation, and coherent excitation promises the best single photon properties, in particular a high quantum state preparation fidelity, as needed for quantum network applications.

This work presents studies on single InAs/InP QDs integrated into mesa structures and excited via pulsed coherent excitation. Coherently driving the biexciton-exciton (XX-X) radiative cascade via triggered two-photon-resonant excitation (TPE) of the XX-state, we demonstrate the on-demand generation of single-photons with high purity in terms of $g^{(2)}(0) \approx 1\%$ and record-high photon-indistinguishabilities of $V_{\text{HOM}}=35(3)\%$ in the telecom C-band. Our work thus represents a notable advancement in the generation of flying qubits for quantum networking in optical fibers.

\begin{figure}[ht]
\centering
\includegraphics[width=0.75 \linewidth]{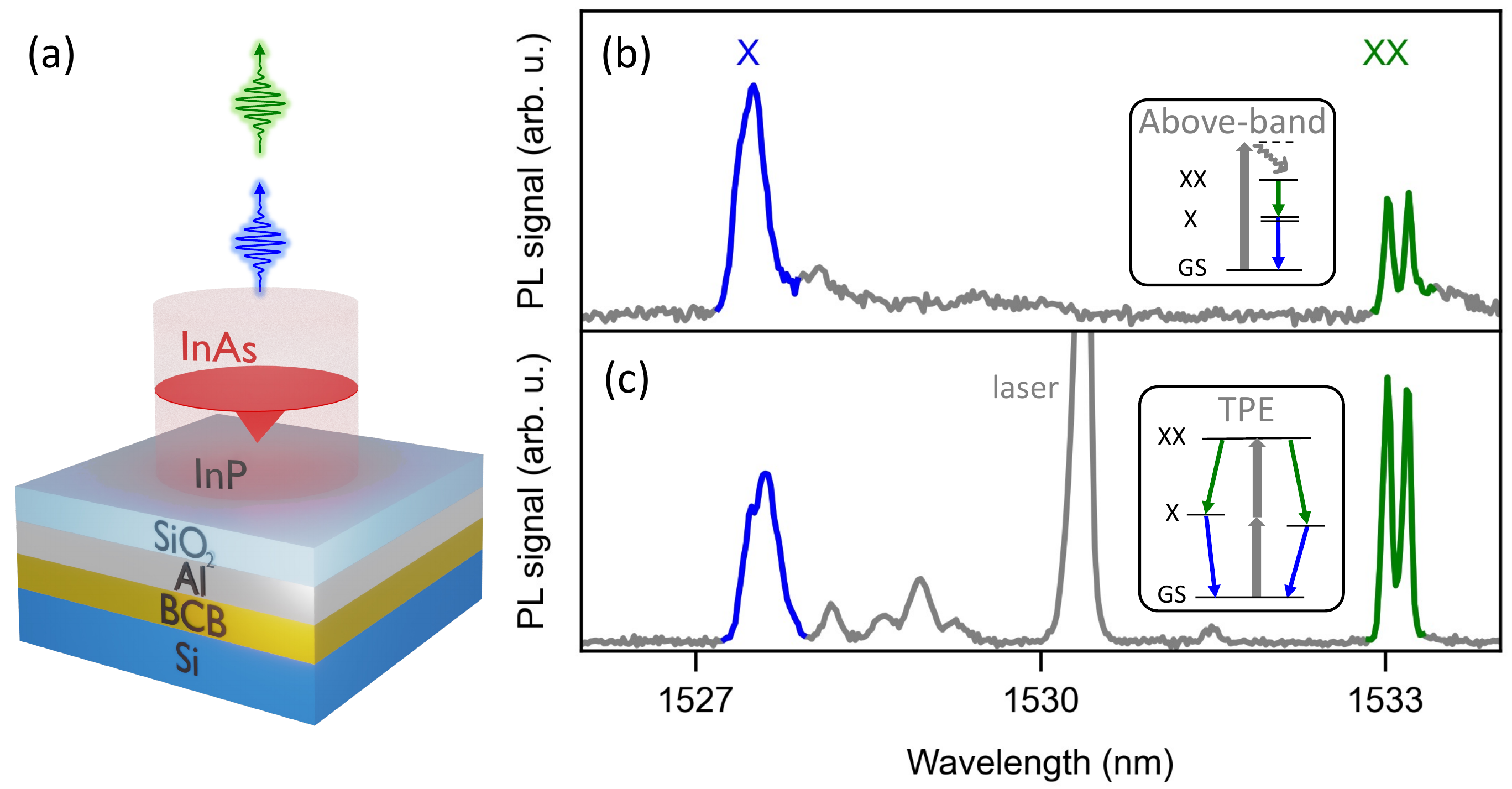}
\caption{(a) Illustration of the Si-compatible QD-mesa structure used in this study: a single InAs QD is embedded in an InP matrix with metallic Al bottom mirror for enhanced photon extraction efficiency. The mesa diameter is $2\,\mu$m. (b) Spectrum of a QD-mesa under above-barrier excitation at saturation power revealing telecom C-band photons originating from the X- (blue) and XX- (green) radiative recombination respectively. (c) Spectrum of the same QD under two-photon resonant (TPE) excitation. The center peak originates from residual scattering of the pump laser. Insets illustrate the excitation scheme used. All spectra are taken at 4$\,$K and the emission is collected via an $\text{NA}=0.7$ aspheric lens.}
\label{figure1}
\end{figure}

\section*{Results and Discussion}
The QD devices (see Fig.~\ref{figure1}(a)) used in this work are based on InAs/InP QDs embedded in mesa structures integrated on a bottom metallic mirror (Aluminum) to enhance the photon extraction efficiency to $\approx 10\%$\cite{holewa2022bright}. Heterogeneous integration on a silicon (Si) carrier thereby allows for a future integration in the complementary metal-oxide-semiconductor (CMOS) platform. See Methods section for details on the device fabrication. Noteworthy in this context, quantum emitter integration on Si has also been demonstrated at O-band wavelengths in the range of 1310-1440$\,$nm \cite{kim2017hybrid,katsumi2022cmos,larocque2023tunable}. Coherent excitation schemes, however, have not yet been implemented on CMOS compatible samples.
\\
The spectrum of a single QD-mesa recorded at 4\,K under above-barrier excitation (continuous wave (CW) laser at 980\,nm) is shown in Fig.~\ref{figure1}(b). Photons originating from the XX-X radiative cascade are observed with wavelengths near 1530\,nm in the telecom C-band. The assignment of the QD states was confirmed via polarization- and excitation-power-dependent photoluminescence measurements as well as photon cross-correlation experiments (see SI, Section \ref{sec:characterization-of-states}). We observe a fine structure splitting (FSS) of the X state of $88(2)\,\mu$eV and a binding energy of the XX-state of $2.9(1)\,\text{meV}$, allowing for TPE of the biexciton state \cite{jayakumar2013deterministic}. Figure~\ref{figure1}(c) depicts a spectrum of the same QD from (b) under pulsed TPE (see Methods for details), revealing the same spectral fingerprint of the QD together with remaining scattered laser light near 1530\,nm. We extract upper bounds for the linewidths under TPE of $119(7)\,\mu \text{eV}$ and $47(2)\,\mu \text{eV}$ FWHM for the X- and XX-transition, respectively (see SI, Section \ref{sec:TPE-linewidth}).
\begin{figure*}[!ht]
\centering
\includegraphics[width=1 \linewidth]{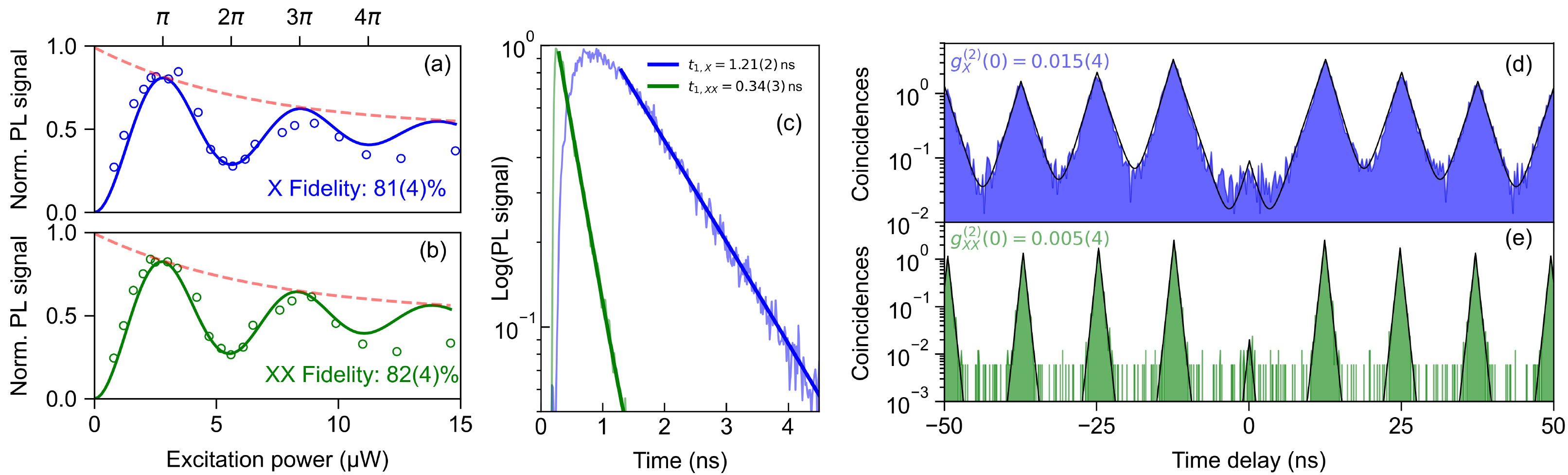}
\caption{(a,b) Tuning the excitation laser power reveals Rabi rotations in the spectrally filtered single-photon detector counts of the X- (blue, upper panel) and XX- (green, lower panel) states up to 4$\pi$, confirming the coherent population of the three-level system. Solid lines correspond to fits and dashed red lines to the extracted envelope function from which preparation fidelities $> 80\%$ are extracted. (c) The decay dynamics illustrate the cascaded emission under pulsed TPE, with the X-photons (blue line) appearing delayed with respect to the XX-photons (green). The XX-state decays $\approx\,$4-times faster than the X-state. (d,e) The second-order auto-correlation measurement confirms the single-photon-nature of X- (blue) and XX- (green) emission.}
\label{figure2}
\end{figure*}

The coherent population of the XX-X cascade under TPE can be confirmed via excitation-power dependent measurements. Figure~\ref{figure2}(a) and (b) display the laser-power-dependent spectrally filtered photoluminescence signal from the X- and XX-state, respectively, detected using a single-photon detector (see Methods for details). Here, clear Rabi rotations are observed up to pulse areas of 4$\pi$ accompanied by a noticeable damping effect. Note that, unlike under strictly resonant excitation, the Rabi frequency scales quadratically with the pulse area under TPE, resulting in equidistant Rabi oscillations as a function of the laser power. Note also that in the presence of damping the pulse area that leads to a complete occupation inversion (the theoretical $\pi$-power) slightly deviates from the power that leads to a maximum in the detected emission. However, here we denote the maximum of the measured Rabi rotation as $\pi$-power for simplicity. 

To estimate the preparation fidelity from the measured Rabi rotations, we follow Refs. \onlinecite{scholl2019resonance,zeuner2021demand} using an exponentially damped oscillation derived by extending an analytical approximation (see Methods and SI, Section \ref{sec:PrepFidelity}, for details). The fit describes the data well up to a pulse area of $4\pi$, while deviating for higher pulse areas. This deviation can be understood by a quenching of the X and XX emission at high powers as the generation of a high number of excess charges at high power favors the formation of trion states. This hypothesis was confirmed by adding a weak above-barrier CW excitation to the pulsed TPE which reduces the X and XX emission signal, while enhancing the trion emission (see SI, Section \ref{sec:TPE-plus-above-barrier} for details). By extrapolating the fitted envelope function of the oscillations (dashed red line) one obtains an estimated preparation fidelity, i.\,e. normalized occupation at $\pi$-power, of $\mathcal{F}_{\text{Prep.}} = 81(4)\%$ and $82(4)\%$ for X and XX respectively which translates into a pair generation fidelity of $66\%$. This observed preparation fidelity compares favorably or is comparable with previously reported values between 49\% and 85\% for QDs at C-band wavelength based on metamorphic buffer layers (see also Table \ref{tab:indistinguishability})\cite{nawrath2019coherence,zeuner2021demand}. Note, that an alternative approach based on photon cross-correlation experiments will be used further below to independently confirm the preparation fidelities obtained above. 

From now on, and unless stated otherwise, experiments were conducted under TPE with a $\pi$-pulse corresponding to $P_{\pi} = 2.5 \,\mu \text{W}$ average excitation power, measured in front of the cryostat window. To gain insights into the dynamics of the three-level system under study, time-resolved measurements were conducted (see Methods for experimental details). In Fig.~\ref{figure2}(c) we observe the typical behavior of the XX-X emission cascade. A fast exponential decay of the XX-state causes the X state to be transiently occupied, followed by a slower exponential decay of the X-state. Applying mono-exponential fits to the time traces, we extract the respective decay constants as
\begin{equation*}
    t_{1,X} = 1.21(2) \,\text{ns} \, , \hspace{3mm}   t_{1,XX} = 0.34(3) \, \text{ns} .
\end{equation*}
Interestingly, the decay of the XX-state is about 4-times faster than the decay of the X-state, a characteristic that was also verified for other QDs on the same sample (see SI, Section \ref{sec:another-dot}). Possible explanations include a specific QD morphology resulting in an increased electron-hole wavefunction overlap for the XX state \cite{bacher1999biexciton,baier2006quantum} or additional decay channels via dark exciton states enabled by intrinsically short spin-flip times as discussed in Ref.$\,$\onlinecite{narvaez2006excitonic}. To generate polarization entangled photon pairs the XX-X emission cascade is typically used. For photons originating from the cascaded decay the indistinguishability is intrinsically limited as the X photons inherit the timing jitter from the XX decay, while the XX photons decay into a spectrally broadened X state. Hence, the achievable indistinguishability is upper bounded by the ratio of the XX- and X-lifetime as $V_{\text{max TPE}}=1/(1+t_{1,XX}/t_{1,X})$\cite{scholl2020crux}. This results in a theoretical limit of $\approx 80\%$ in our case, which significantly exceeds the value of $67\%$ resulting from the often assumed case of $t_{1,X} = 2 \times t_{1,XX}$. Thus, the InAs/InP QDs investigated here, which exhibit a larger $t_{1,X}/t_{1,XX}$ ratio, are promising candidates for combining entanglement generation with high photon indistinguishability. Technologically more involved approaches in this direction aim to exploit microcavities supporting an asymmetric Purcell enhancement for maximizing this lifetime-ratio \cite{bauch2023demand}.

Next, we investigated the purity of the single photons in terms of the second-order auto-correlation function $g^{(2)}(\tau)$ via Hanbury-Brown and Twiss type experiments. The resulting $g^{(2)}(\tau)$-histograms for X- and XX-photons are presented in Fig.~\ref{figure2}(d) and (e), respectively. The strong suppression of coincidences at zero time delay confirms the emission of single photons. The experimental data are approximated by a fit function corresponding to a sum of two-sided exponential decays and accounting for the noticeable blinking effect on short timescales, following Ref.~\onlinecite{holewa2023scalable} (see SI, Section \ref{sec:HBT-fit} for details), while no temporal deconvolution or background subtraction was applied. The fit yields antibunching values of 
\begin{equation*}
    g_{X}^{(2)}(0) = 0.015(4) \, , \hspace{3mm} g_{XX}^{(2)}(0) = 0.005(4),
\end{equation*}
where the errors have been determined from the fit residuals. The decay times obtained from the fit are $1.44(1)\,\text{ns}$ and $0.36(1)\,\text{ns}$ for the X- and XX-state, respectively, in good agreement with the time-resolved photoluminescence measurements. We further extract the timescale of the blinking as ${\tau_{\mathrm{blink}} \approx 17(1)\,\text{ns}}$, which can be caused by either charging events or spectral wandering due to fluctuation in the QD's charge environment.

\begin{figure*}[!ht]
\centering
\includegraphics[width=1 \linewidth]{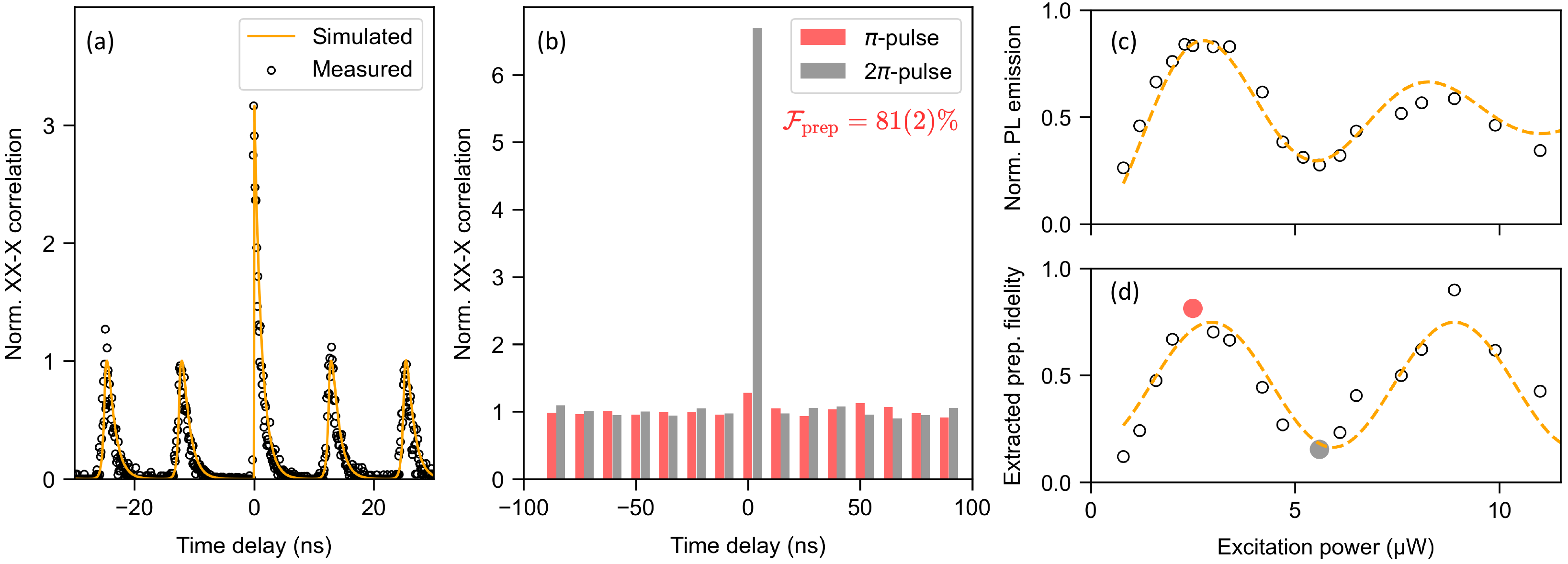}
\caption{Analysis of the preparation fidelity via XX-X photon cross-correlation experiments: (a) Measured cross-correlation histogram after polarization and blinking correction (grey dots), matching the Monte Carlo simulation (orange line) of expected polarization-filtered XX-X correlation for a QD with a DOP of 33$\%$ and a preparation fidelity of 81$\%$. (b) Comparison of polarization-corrected and integrated coincidences at $\pi$-power (red) and at $2\pi$-power (grey, shifted for clarity) shows difference in preparation fidelity. (c) Summed XX and X emission intensity oscillation as a function of excitation power with Rabi fit from Fig.$\,$\ref{figure2}(a) (orange line) for comparison. (d) Preparation fidelity $\mathcal{F}_{\text{prep}}$ extracted from cross-correlation histograms as a function of excitation power. Red and grey circles indicate the data sets shown in panel (b) (orange line is a guide to the eye).}
\label{figure4}
\end{figure*}

Having confirmed the coherent population of the XX-X cascade via TPE by observing Rabi rotations in Fig.$\,$\ref{figure2}(a,b), revealing a fidelity for preparing the XX- (X-) state of $82(4)\%$ ($81(4)\%$), we applied another independent method for extracting the preparation fidelity to verify our results. For this purpose, photon cross-correlation experiments between photons emitted from the XX- and X-state were carried out \cite{neuwirth2022multipair} (see Methods and SI, Section \ref{sec:PrepFidelity} for details). Fig.$\,$\ref{figure4}(a) depicts the photon cross-correlation histogram recorded at $\pi$-pulse excitation (black circles) together with a Monte Carlo simulation (orange line) accounting for our experimental conditions (including setup imperfections). The experimental data, obtained under TPE at $\pi$-power and corrected for polarization- and blinking effects as mentioned above, is consistent with a preparation fidelity of 81\%. The simulations are in very good agreement and also reproduce fine details, such as the correct area ratios, the asymmetry of the side peaks caused by the faster XX decay, and the strong asymmetry in the center peak caused by the cascaded decay (see SI, Section \ref{sec:simulation-cross-correlation} for details on simulation). The comparison of the integrated coincidence-peak areas depicted in Fig.$\,$\ref{figure4}(b) yields a maximum preparation fidelity of $81(2)\%$ at $\pi$-power (red bars) in perfect agreement with the value determined from the Rabi rotation in the emission intensities in Fig.$\,$\ref{figure2}(a,b). The error is determined from the standard deviation of the non-center peak areas. At $2\pi$-power (grey bars) the preparation fidelity reduces to $14(2)\%$. Complementary data under off-resonant excitation (see SI, Section \ref{sec:PrepFidelity}) yielded a preparation fidelity of $44(2)\%$), clearly confirming the positive impact of coherent excitation.
By extracting the preparation fidelity from the cross-correlation experiments as a function of the excitation power (cf. Fig.$\,$\ref{figure4}(d)), oscillations of the preparation fidelity can be observed which are in phase with the Rabi rotations of the emission (cf. Fig.$\,$\ref{figure4}(c)). Interestingly, the fidelity oscillations do not exhibit the same damping observed in the Rabi rotations.

\begin{figure}[!ht]
\centering
\includegraphics[width=0.75 \linewidth]{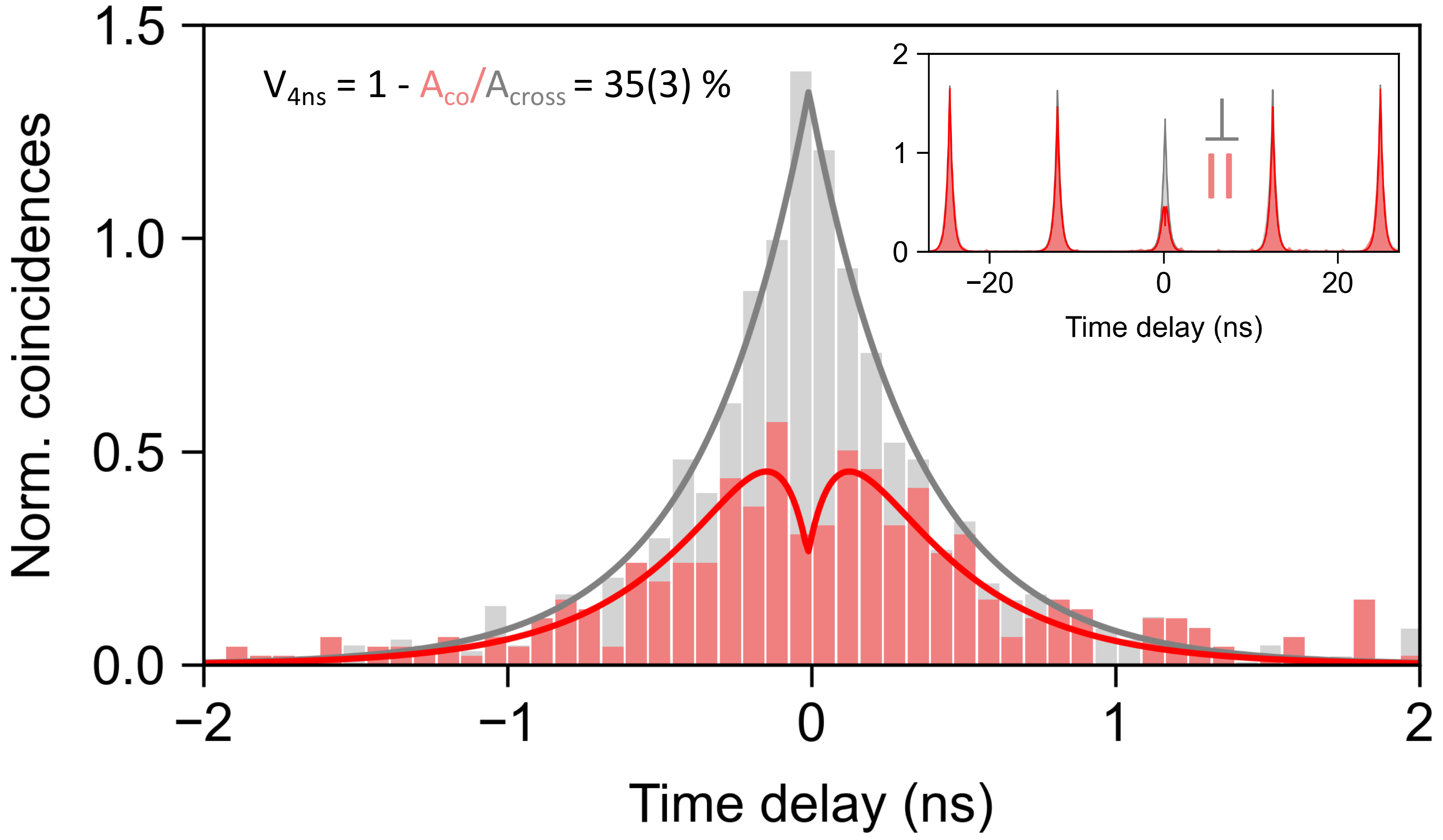}
\caption{HOM-type two-photon interference experiments of photons emitted by the XX-state for co- (red) and cross- (grey) polarized measurement configuration. Photon coalescence described by the Hong-Ou-Mandel effect is confirmed by the clear suppression of coincidences in the central peak for co-polarized photons. We observe record-high photon-indistinguishabilities of $V_{\text{raw}}=29(2)\%$ and $V_{\text{4\,ns}}=35(3)\%$ by integrating the raw experimental data over the full 12.5\,ns- and a 4\,ns-wide delay-window, respectively.
The inset shows the same data for a broader range of arrival time delays.}
\label{figure3}
\end{figure}

Finally, we explored the indistinguishability of the XX-photons emitted by the coherently driven three-level system via Hong-Ou-Mandel (HOM) type two-photon interference experiments (see Methods for details). The resulting experimental data for the co- and cross-polarized measurement configuration is presented in Fig.$\,$\ref{figure3} as a close-up highlighting the zero-delay peak. The reduction of coincidences due to two-photon interference is clearly visible. Additionally, the inset depicts the histograms for larger arrival time delays. Note, that the ratios of the coincidence side-peak areas deviate from the expected behavior as they are masked by the blinking effect. This however does not affect the following analysis, which is solely based on the integrated zero-delay peak areas for co- and cross-polarized measurements (without applying the fit model). 

The two-photon interference visibility is extracted from our measurement data via $V_{\text{HOM}}=1 - (A_{\text{co}}/A_{\text{cross}}),$ with the integrated areas $A_{\text{co}}$ and $A_{\text{cross}}$ of the co- and cross-polarized case, respectively. Integrating over the entire zero-delay time window, i.$\,$e. $-6.25\,$ns to $+6.25\,\text{ns}$ (corresponding to the laser repetition rate of 80$\,\text{MHz}$), yields $V_{\text{raw}}=29(2)\%$, with the error inferred from the distribution of the integrated detection events of the non-interfering side-peaks. This photon-indistinguishability readily exceeds all previous reports on pulsed HOM experiments of telecom C-band photons directly generated via QDs (cf. Table \ref{tab:indistinguishability}), including pioneering work by Nawrath et al.~\cite{nawrath2021resonance,nawrath2023bright} as well as our most recent work in Ref. \onlinecite{holewa2023scalable}. 

\begin{table*}[ht]
\centering
\caption{Comparison of reported pulsed HOM visibilities $V_{\text{HOM}}$ and preparation fidelities $\mathcal{F}_{\text{prep}}$ for single photons emitted by QDs in the telecom C-Band (MB: Metamorphic buffer layer, CBG: Circular Bragg grating, R: resonant, TPE: two-photon-resonant, QR: quasi-resonant). The subscript of the visibility $V$ refers to the coincidence histogram integration window when evaluating the HOM measurement.}
\label{tab:indistinguishability}
\begin{tabular}{l|l|l|l|l|l}
QD Type & Device & Excitation  & pulsed $V_{\text{HOM}}$ & $\mathcal{F}_{\text{prep}}$ & Reference \\ \hline \hline
  InAs/InP & Mesa & TPE & $V_{\text{4\,ns}}=35(3)\%$ & $81(2)\%$ & This work \\
   InAs/InP & Mesa & TPE & $V_{\text{raw}}=29(2)\%$ & $81(2)\%$  & This work \\
    InAs/InP &  CBG & QR & $V_{\text{4\,ns}}=19(3)\%$ & - & Holewa et al. 2023\cite{holewa2023scalable} \\
     InAs/InGaAs + MB  &  CBG & QR & $V_{\text{6\,ns}} \approx 10\%$ & - & Nawrath et al. 2023\cite{nawrath2023bright} \\
     InAs/InGaAs + MB &  planar & TPE & - & $84(3)\%$ & Zeuner et al. 2021\cite{zeuner2021demand}  \\
     InAs/InGaAs + MB &  planar & R & $V_{\text{4\,ns}}=14(2)\%$ &  & Nawrath et al. 2021\cite{nawrath2021resonance}  \\
      InAs/InGaAs + MB  &  planar & R & - & $49(6)\%$ & Nawrath et al. 2019\cite{nawrath2019coherence}  \\ \hline 
\end{tabular}
\end{table*}

Note, that in these earlier reports, typically photons with a temporal separation of 4$\,$ns were interfered. Thus, for the sake of comparison, we also evaluate the two-photon interference visibility for a 4$\,$ns-wide integration window as {$V_{\text{4\,ns}}=35(3)\%$}. Importantly, this integration windows covers over $99\%$ of all coincidences detected in the zero-delay peak. Thus, the increase of $V_{\text{4\,ns}}$ compared to $V_{\text{raw}}$ results from the improved signal-to-noise ratio (due to reduced noise contributions by dark counts) rather than discarding signal events. In addition, by interfering photons with a temporal separation of $12.5\,\text{ns}$, the QD-excitonic states investigated in our work are potentially subject to larger dephasing compared to previous studies that used $\Delta t=4\,$ns \cite{thoma2016exploring}.

While the photon-indistinguishability observed in our work represents a substantial improvement compared to previous reports on QD photons emitted directly at C-band wavelengths, it is not yet competitive with up-converted photons from QDs emitting at lower wavelengths \cite{da2022pure}. As the achieved indistinguishability is well below the limit of 80$\%$ set by the ratio of the measured radiative lifetimes \cite{scholl2020crux}, there are still other limiting factors. The main contribution currently limiting the photon-indistinguishability is suspected to be charge fluctuations in the QD environment, as evidenced by a considerable blinking effect in the auto-correlation measurement, the presence of a predominant trion state under above-barrier excitation, as well as the influence of additional weak CW above-barrier excitation during TPE (cf. SI, Section \ref{sec:TPE-plus-above-barrier}), which reduces the coherence time of the QD under study. We anticipate two routes to overcome these limitations. Firstly, implementing measures to influence or tailor the electric-field environment of the quantum emitters, using charge stabilization via electrical gates \cite{somaschi2016near,zhai2020low} or interface passivation \cite{tomm2021bright}. Secondly, employing droplet-epitaxy \cite{Holewa2022}, which is known to increase the coherence time \cite{anderson2021coherence}, rather than self-assembled Stranski-Krastanov growth of QDs as used in our work. Both routes have independently shown to improve the coherence properties of specific QD systems.

Note for completeness that by performing a fit which includes the central dip (solid lines in Fig.$\,$\ref{figure3}) and accounting for the blinking effect (see SI, Section \ref{sec:HOM-fit} for details), one can extract the post-selected HOM visibility which amounts to $73(6)\%$ and thus compares well with previous works performing CW HOM experiments on similar QDs \cite{anderson2020quantum,zeuner2021demand}. One should emphasize, however, that the pulsed HOM result presented above is the most relevant one for applications in quantum information, as it describes the interference of the full photon wave packets.
The coherent on-demand generation of indistinguishable photons via TPE at telecom C-band wavelength achieved in our work represents a significant advance towards applications of quantum information and long-haul quantum communication in optical fibers. 
Yet, the most suitable excitation scheme for implementations of quantum communication needs to be carefully selected depending on the specific use-case.
For defining time bins and synchronizing sender and receiver units, pulsed excitation is needed, while coherent excitation provides the highest photon indistinguishability. Among all known approaches for pumping QD emitters, coherently driving the XX-X cascade via TPE promises the highest purity in terms of $g^{(2)}(0)$, as it reduces the re-excitation probability \cite{hanschke2018quantum}. In addition, the spectrally detuned excitation laser allows for straightforward spectral filtering of the emitted photons and for the generation of polarization-entangled photon-pairs \cite{huber2018semiconductor}. While the maximum indistinguishability is limited in standard TPE \cite{scholl2020crux}, it can be further boosted by stimulating the XX-decay channel to distill highly indistinguishable X-photons \cite{wei2022tailoring}.

Another important aspect in quantum communication concerns whether the emitted photons form a pure or a mixed state in the photon number basis, which strongly depends on the excitation scheme \cite{bozzio2022enhancing}. For most quantum communication protocols the absence of photon number coherence (PNC) is required to avoid side-channel attacks and maintain security. As recently demonstrated for QD-generated photons at shorter wavelengths, standard TPE does not result in a significant degree of PNC in the emitted photon states, which can however be recovered and tuned by stimulating the XX-decay channel with a second laser pulse \cite{karli2023controlling}.

\section*{Conclusion}
In summary, we demonstrated the pulsed coherent excitation of InAs/InP QDs emitting photons at telecom C-band wavelengths with unprecedented photon-indistinguishability. Using triggered TPE of the XX-X radiative cascade, we coherently populate the three-level system, which is confirmed by the observation of Rabi rotations up to 4$\pi$ consistently yielding preparation fidelities exceeding $80\%$, as confirmed by two independent experimental approaches. The observed two-photon-interference visibility of up to 35(3)$\%$ clearly surpasses previous results obtained for triggered photons in the telecom C-band directly emitted by QDs. The type of QDs studied in our work exhibits a large ratio of exciton-to-biexciton lifetime, making them promising candidates for the generation of polarization-entangled photon-pairs with high photon-indistinguishability at C-band wavelength. The demonstrated on-demand generation of QD photons with record-high indistinguishability at wavelengths compatible with existing fiber infrastructure presents a significant step towards scalable quantum networks.

\subsubsection*{Note Added in Proof}
During the review process of our work, we became aware of related work by Joos et al. \cite{joos2023triggered}, reporting experiments on the generation of indistinguishable QD telecom C-band photons using phonon-assisted excitation as well as the swing-up of quantum emitter population scheme \cite{bracht2021swing}.

\section*{Methods}
\subsection*{QD Device}
The QD devices used in this work are based on InAs/InP QDs obtained via self-assembled Stranski-Krastanow growth in the near-critical regime via metalorganic vapor phase epitaxy (MOVPE) \cite{berdnikov2023fine}. To enhance the photon extraction efficiency, mesa structures are combined with a bottom metallic mirror (Aluminum), while bonding via benzocyclobutene (BCB) to the Si-carrier allows future integration in the CMOS platform. For details on the device fabrication and basic characterization we refer the interested reader to the supplementary information (SI), Section \ref{sec:sample-fabrication} and \ref{sec:characterization-of-states}, as well as Ref. \onlinecite{holewa2022bright}.

\subsection*{Quantum Optics Experiments}
The QD devices are investigated by means of low-temperature (4\,K) micro-photoluminescence spectroscopy. Samples are mounted inside a closed-cycle cryostat integrated in a cryooptical table (model attoDRY800, attocube systems AG). An aspheric lens $\text{NA}=0.7$ inside the cryostat is used to collect the QD emission. For optically pumping single QDs, we use a 980\,nm CW laser or a spectrally shaped wavelength-tunable 2.5\,ps pulsed laser system (picoEmerald, APE GmbH) with 80\,MHz repetition rate. For TPE the picosecond laser is combined with a folded 4f pulse-shaper, resulting in a spectral width of $\approx 0.5\,\text{nm}$ full-width half-maximum (FWHM) and a stretched pulse duration of 10$\,$ps FWHM.
Photons originating from the X- or XX-state are spectrally analyzed using a grating spectrometer (900\,lines/mm) with an attached InGaAs array detector. For time-correlated single-photon counting experiments, the X- and XX-photons are spectrally filtered using a tunable fiber-bandpass filter of 0.4$\,$nm bandwidth in combination with polarization suppression of the scattered excitation laser. Time-resolved measurements were conducted using superconducting nanowire single photon detector (SNSPD) (SingleQuantum EOS, SingleQuantum BV) in combination with time-tagging electronics (quTag, quTools GmbH) with a combined timing resolution of $\approx 50\,\text{ps}$ FWHM.

To measure the purity we performed Hanbury-Brown- and Twiss auto-correlation experiments by passing the spectrally filtered single photons through a 50:50 fiber beamsplitter and registering the arrival time difference between the detectors at the two outputs.

The photon-indistinguishability is measured in HOM-type experiments using a Mach-Zehnder interferometer with 12.5$\,$ns delay, compensating the photon delay imposed by the $80\,$MHz excitation laser. Here, interfering indistinguishable photons cause coalescence in the two interferometer outputs ports, thus reducing the number of detected coincidences. Additionally, the polarization state of the interfering photons is controlled in both arms of the interferometer, allowing for measurements in co- and cross-polarised configuration, whereas the latter serves as a reference to extract the two-photon-interference visibility $V$. Here, the visibility was extracted by integrating over all coincidences within one excitation cycle for the co- and cross-polarized measurement, as well as by selecting a realistic 4$\,$ns time window. For details concerning the data analysis of the time-resolved experiments we refer to SI, Section \ref{sec:HBT-fit} and \ref{sec:HOM-fit}.

\subsection*{Estimating Preparation Fidelities}
Different approaches can be used to estimate the preparation fidelity. From the Rabi rotations in emission intensity vs. excitation-power this is possible, e.g., by comparing the occupation at $\pi-$ and $2\pi$-power\cite{muller2014demand}, fitting exponentially damped $\sin^2$-functions\cite{reindl2018all}, solving rate equation models numerically \cite{wang2005decoherence,nawrath2019coherence}, or by modeling the full system via polaron-transformed open system master equations \cite{mccutcheon2011general} or correlation expansion approaches to include phonons \cite{krugel2005role}. To independently confirm the preparation fidelities obtained in our work, we used two different experimental approaches.

In the first approach, used in Figure~\ref{figure2}(a) and (b), we model the experimental Rabi data following Refs. \onlinecite{scholl2019resonance,zeuner2021demand}, fitting the experimental Rabi data using an exponentially damped oscillation derived by extending an analytical approximation for CW Rabi oscillations to Rabi rotations in the pulsed regime.

The second approach, used in Figure~\ref{figure4}(b), is based on photon cross-correlation experiments between photons emitted from the XX- and X-state \cite{neuwirth2022multipair}. Comparing the integrated coincidences originating from the same cascade $A_{\text{same}}$ with those from different cascades $A_{\text{different}}$, the preparation fidelity can in principle be extracted as $\mathcal{F}_{\text{prep}}=A_{\text{different}}/A_{\text{same}}$ \cite{wang2019demand}. Importantly, the experimental data had to be corrected to account for a partial polarization of the QD emission, the polarization-selective detection in our setup, and the blinking effect discussed above.

For details on the analysis and the required corrections we refer to the SI, Section \ref{sec:PrepFidelity}.

\subsubsection*{Author Contributions}
The concept of the QD device was suggested by E.S. and realized under her supervision at DTU. Epitaxy and nano-fabrication was performed by P.H. with support of A.S. and K.Y.. The sample design was supported by simulations performed by N.G.. Initial optical characterization of the sample was performed in Wrocław by P.H. with support of E.Z., M.W, and A.M. and supervised by M.S.. A.H. contributed to the data analysis.
The experiments, simulations, and data analysis in this work were conducted in Berlin by D.A.V. under the supervision of T.H. and M.v.H.. The manuscript was written by D.A.V., T.H., and M.v.H. with input from all authors. T.H. supervised all efforts in this project.

\subsubsection*{Acknowledgement}
The authors gratefully acknowledge helpful discussions with Yusuf Karli, Florian Kappe, Thomas Bracht, and Doris Reiter.

\subsubsection*{Funding Sources}
We acknowledge financial support by the German Federal Ministry of Education and Research (BMBF) via the project “QuSecure” (Grant~No.~13N14876) within the funding program Photonic Research Germany, the BMBF joint project “tubLAN Q.0” (Grant~No.~16KISQ087K), the Einstein Foundation via the Einstein Research Unit “Quantum Devices”, the Danish National Research Foundation via the Research Centers of Excellence NanoPhoton (DNRF147) and the Center for Macroscopic Quantum States bigQ (DNRF142).
P.\,H. was funded by the Polish National Science Center within the Etiuda~8 scholarship (Grant~No.~2020/36/T/ST5/00511).
N.\,G. acknowledges support from the European Research Council (ERC-CoG “UNITY”, Grant No. 865230), and from the Independent Research Fund Denmark (Grant No. DFF-9041-00046B).

\putbib
\end{bibunit}
\newpage

\begin{bibunit}
\setcounter{figure}{0}
\setcounter{table}{0}
\renewcommand\thefigure{S\arabic{figure}}
\renewcommand\thetable{S\arabic{table}}
\captionsetup[figure]{name=Figure}
\captionsetup[table]{name=Table}

\section*{Supporting Information for: On-Demand Generation of Indistinguishable Photons in the Telecom C-Band Using Quantum Dot Devices}
\section{Quantum Dot Characterization}
\subsection{Sample Fabrication}
\label{sec:sample-fabrication}
The QD structure was grown on (001)-oriented InP substrate by metal-organic vapour phase epitaxy (MOVPE). The array of low surface density ($\sim 2.8 \times 10^{9}$cm$^{ - 2}$) InAs QDs was placed in the centre of a 488 nm thick InP layer grown on a 200 nm-thick InGaAs sacrificial layer lattice-matched to InP. Subsequently, 100 nm thick SiO$_2$ followed by a 100-nm-thick metallic reflector were deposited onto InP surface, and then bonded to a Si substrate using benzocyclobutene (BCB). Finally, the InP substrate and sacrificial layer were removed. To fabricate mesas, we used electron beam lithography followed by inductively coupled plasma-reactive ion etching. For more details we refer to Ref.~\onlinecite{holewa2022bright}. 

\subsection{Identification of Excitonic Complexes}
\label{sec:characterization-of-states}
This section provides additional experimental data supporting the assignment of the excitonic states of the QD investigated in the main article. The measurements shown in Supplemental Information (SI) Figure \ref{fig:SI_identification} were performed under above-barrier continuous wave (CW) excitation using a 980$\,$nm laser diode. The assignment of emission lines to the biexciton (XX) and exciton (X) state of the QD is confirmed by polarization- (cf. Figure \ref{fig:SI_identification}(a)) and excitation-power dependent (cf. Figure \ref{fig:SI_identification}(b)) photoluminescence measurements collecting only emission from the single QD device (µPL). The identification of the excitonic complexes is further supported by XX-X photon cross-correlation measurements (cf. Figure \ref{fig:SI_identification}(c)). The polarization-resolved µPL spectra reveal a fine-structure splitting (FSS) of the X state of $(88\pm 2)\,\mu$eV in line with Ref. \onlinecite{holewa2022bright}. Moreover, analyzing the total X and XX emission intensity as a function of the detection polarization, a degree of linear polarization (DOP) of 33$\%$ can be inferred from the oscillation contrast (cf. Figure \ref{fig:SI_identification}(a), Inset). The excitation-power dependent µPL, presented in double-logarithmic scaling, reveals a saturation of the X emission (blue line), a super-linear increase of the XX emission (green line) at high power and a monotonic behavior of the trion emission (black line). Not least, the cascaded emission of XX- and X-photons is confirmed by the asymmetric photon cross-correlation signature in Figure \ref{fig:SI_identification}(c).

\begin{figure}[ht]
    \centering    \includegraphics[width=\textwidth]{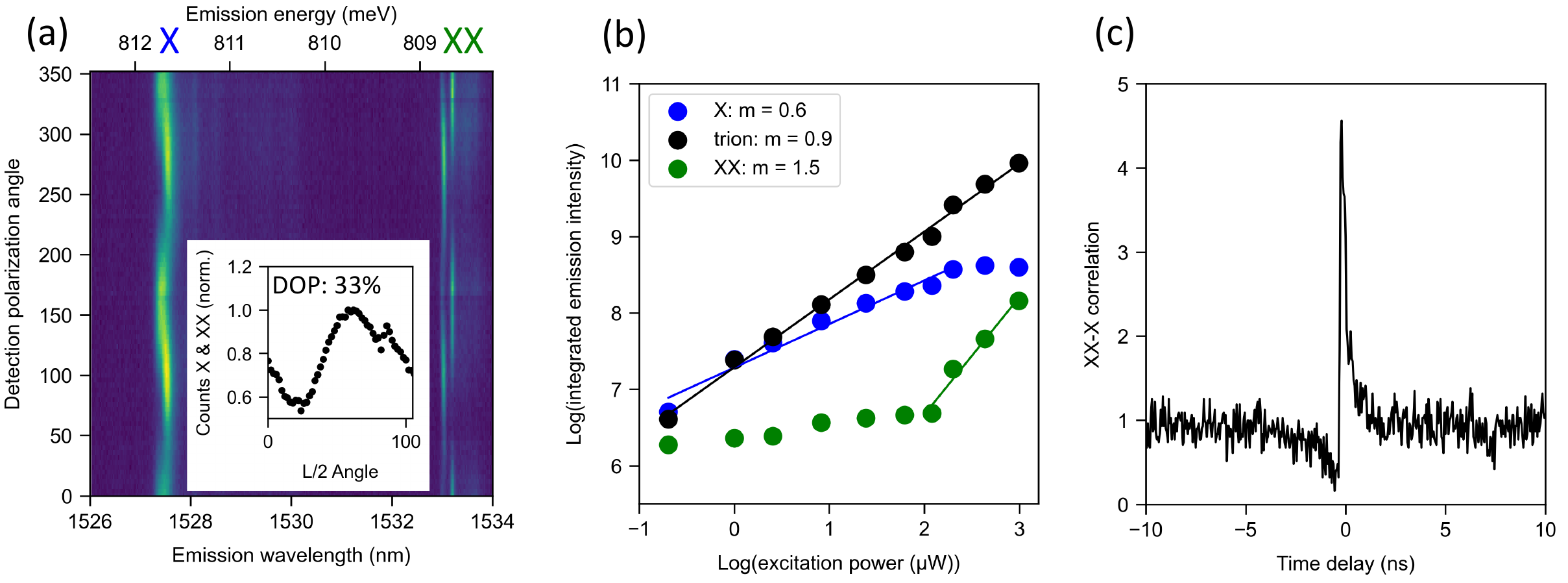}
    \caption{Identification of excitonic complexes under above-barrier excitation: (a) Micro-photoluminescence spectra as a function of the angle between a $\lambda/2$-waveplate and an analyzing polarizer, revealing a fine-structure splitting (FSS) of the biexciton (XX) and exciton (X) emission doublets. Inset: Summed integrated intensity of XX and X emission for extraction of the degree of linear polarization (DOP). (b) Integrated emission intensities of XX, X, and trion-state as a function of the excitation power. (c) Photon cross-correlation histogram using the XX-photons as 'start' and X-photons as 'stop'.} \label{fig:SI_identification}
\end{figure}

\subsection{Linewidth under Two-Photon-Excitation}
\label{sec:TPE-linewidth}
To extract the QD emission linewidths under TPE, we perform Gaussian fits of the X and XX emission line (see Figure \ref{fig:SI_white-light} (a,b)). The FSS is more clearly resolved for the XX emission, as the four-particle state is shielded better against Coulomb forces in the QD environment, leading to a spectral broadening of the X emission lines even at lowest excitation power. In this case, a sum of two overlapping Gaussian fits is used to extract the individual linewidths as well as the FSS. This yields values for the FSS under TPE of $83(6)\,\mu$eV for the X and $91(2)\,\mu$eV for the XX state, in agreement with the results obtained under above-barrier characterization  \cite{holewa2022bright} and the value extracted from Figure \ref{fig:SI_identification}. We obtain linewidths (full-width at half-maximum) of the individual FSS components for the XX- and X-emission of $47(2)\,\upmu \text{eV}$ and $119(7)\,\upmu \text{eV}$, respectively, and a binding energy of the XX-state of $2.9(1)\,\text{meV}$.

\subsection{Effect of weak above-barrier Support}
\label{sec:TPE-plus-above-barrier}
\begin{figure}[ht]
    \centering
    \includegraphics[width=\textwidth]{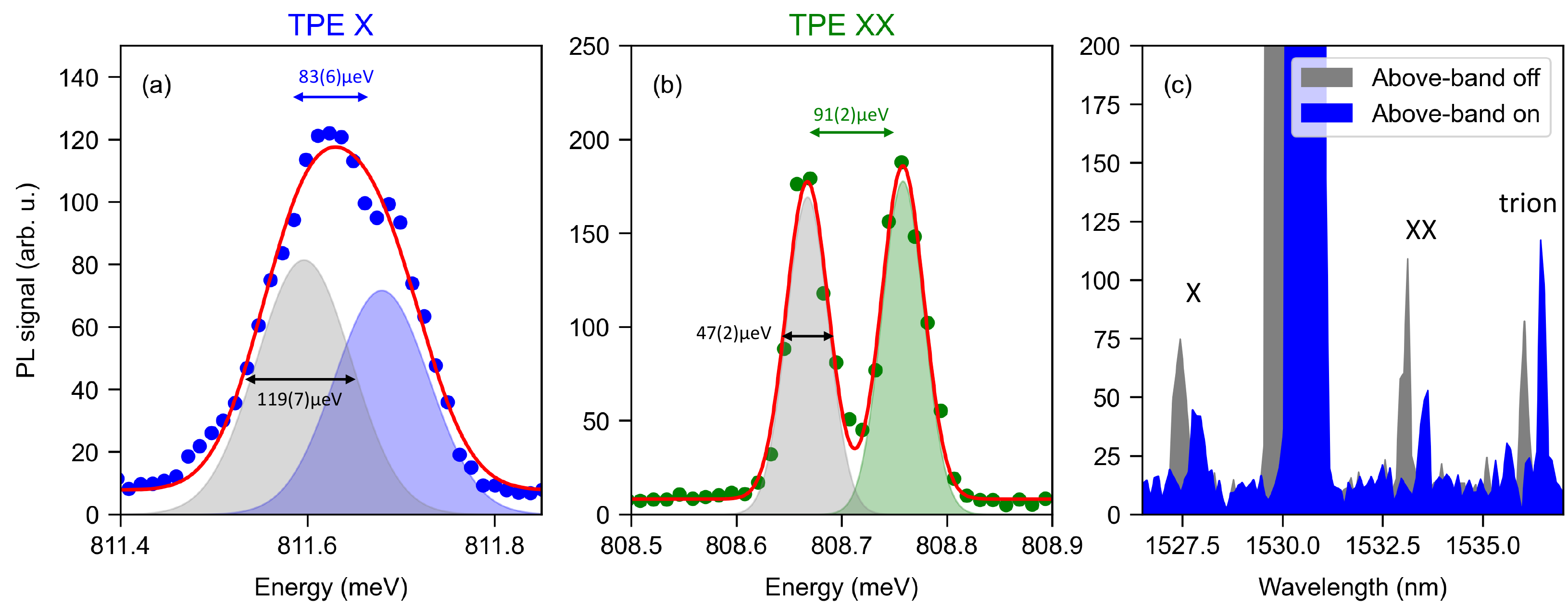}
    \caption{(a,b) Double-Gaussian fits of the X and XX emission under TPE are used to extract the linewidths, fine-structure splitting (FSS), and the XX binding energy. (c) Emission of the QD under TPE without (black) and with (blue) additional weak above-barrier CW excitation, respectively. Data in blue are shifted for better comparison. The partial quenching of the X and XX emission in combination with the enhanced trion signal are indicative for a larger amount of excess charges.}
    \label{fig:SI_white-light}
\end{figure}

As stated in the main article, the Rabi rotations under two-photon-excitation (TPE) cannot be observed at high excitation powers, as the emission of the exciton (X) and biexciton (XX) is quenched, which also results in deviations of the measured Rabi rotation signal from the theoretical fit at higher pulse areas $>4\pi$ (cf. main article, Figure 2(a,b)). This behavior is explained as follows: At high pulse areas, the laser generates a large amount of excess charge carriers, which results in dominant emission via the trion state, as an additional charge is always available. This interpretation is confirmed by the observation that introducing additional weak above-barrier excitation(around 780$\,$nm) CW light (not enough to result in noticeable QD emission) to the TPE pulses results in an enhancement of the trion emission, but a quenching of the XX-X emission (see Figure \ref{fig:SI_white-light}(c)). At high excitation powers this effect can occur without additional CW above-barrier support, explaining the reduced X and XX emission intensity at high pulse areas.

\begin{figure}[ht]
    \centering
    \includegraphics[width=0.75\textwidth]{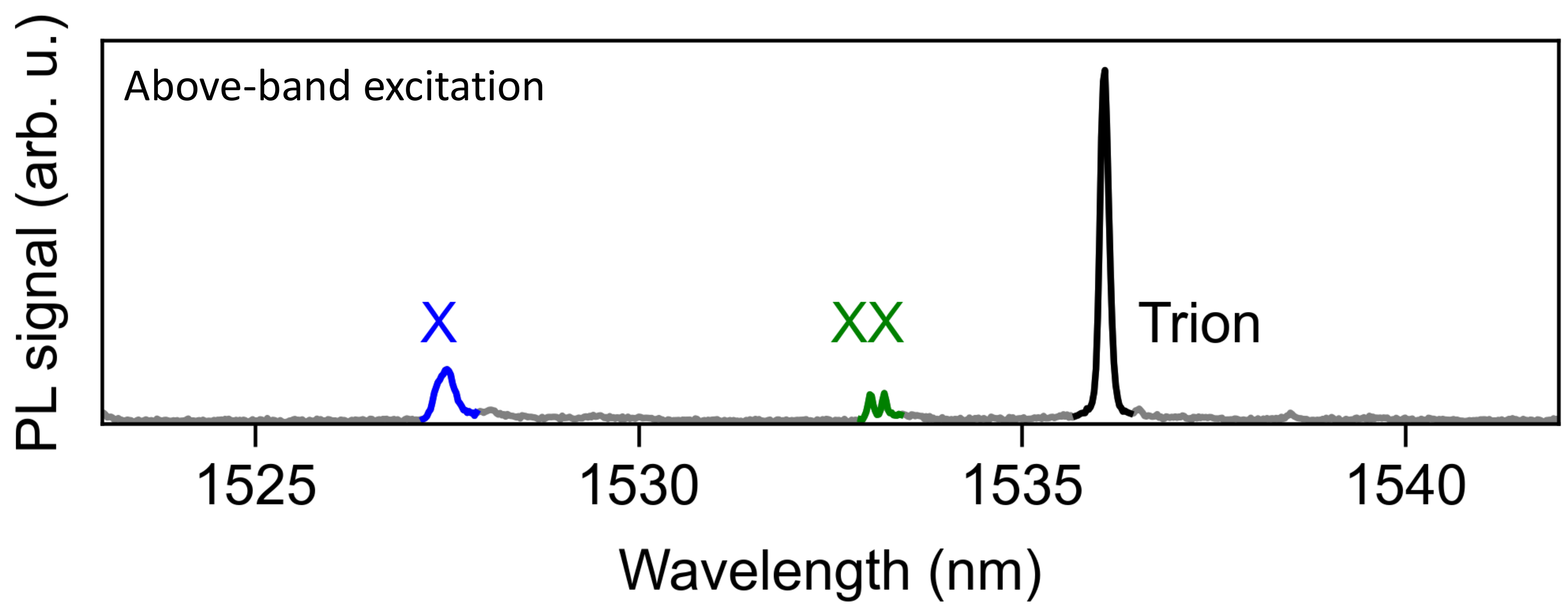}
    \caption{Photoluminescence spectrum under above-barrier excitation with a CW laser at 980$\,$nm. Same spectrum as Figure 1(b) of the main text, but including the dominant trion contribution here.}
    \label{fig:SI_trion}
\end{figure}

The effect of excess charge carriers is further enhanced by the presence of an unintentional background doping in the sample investigated in our work. The dominant emission of the trion state, not shown in Figure 1(b) in the main article, under above-barrier excitation already indicates the presence of a large number of intrinsic charge carriers (cf. Figure \ref{fig:SI_trion}). 
Interestingly, it has previously also been reported to be beneficial to add additional weak CW above-barrier excitation, which helped to increase the Rabi rotation amplitude under TPE by saturating the charge environment in Ref. \onlinecite{zeuner2021demand}. This underlines the sample specific character of effects like intrinsic background doping. 

\subsection{Comparison of Lifetime Measurements under TPE and above-barrier Excitation}
\label{sec:lifetime-comparison}
\begin{figure}[ht]
    \centering
    \includegraphics[width=\textwidth]{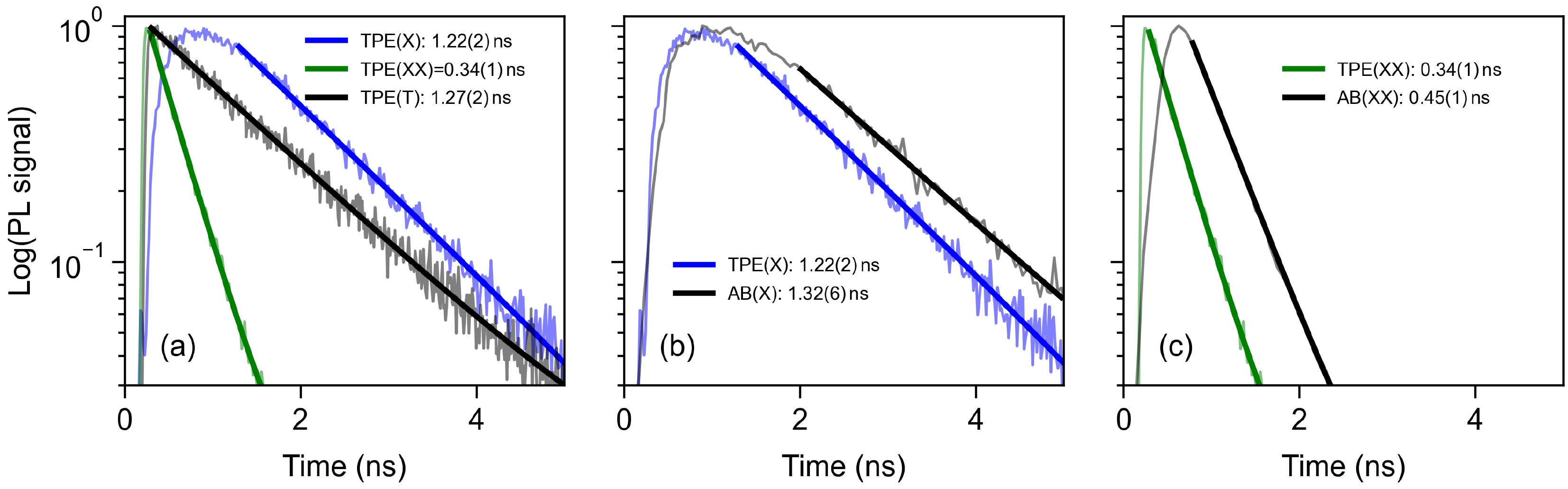}
    \caption{(a) Lifetime measurements for the X-, XX-, and trion- (T) state from the QD studied in the main article. (b) and (c) Comparative study of the XX- and X-lifetimes under pulsed TPE and pulsed above-barrier (AB) excitation.}
    \label{fig:SI_lifetime}
\end{figure}

This section provides additional data of the lifetime measurement for the trion state from the QD studied in the main article (see Figure \ref{fig:SI_lifetime}(a)), including lifetimes of XX and X-state) and a quantitative comparison of the lifetime measurements for the XX- and X-state under TPE and above-barrier (AB) excitation, respectively (cf. Figure \ref{fig:SI_lifetime}(b,c)). The experimental results indicate a slightly reduced decay time under direct coherent excitation via TPE.

\subsection{Two-Photon Excitation of another QD}
\label{sec:another-dot}
\begin{figure}[ht]
    \centering
    \includegraphics[width=\textwidth]{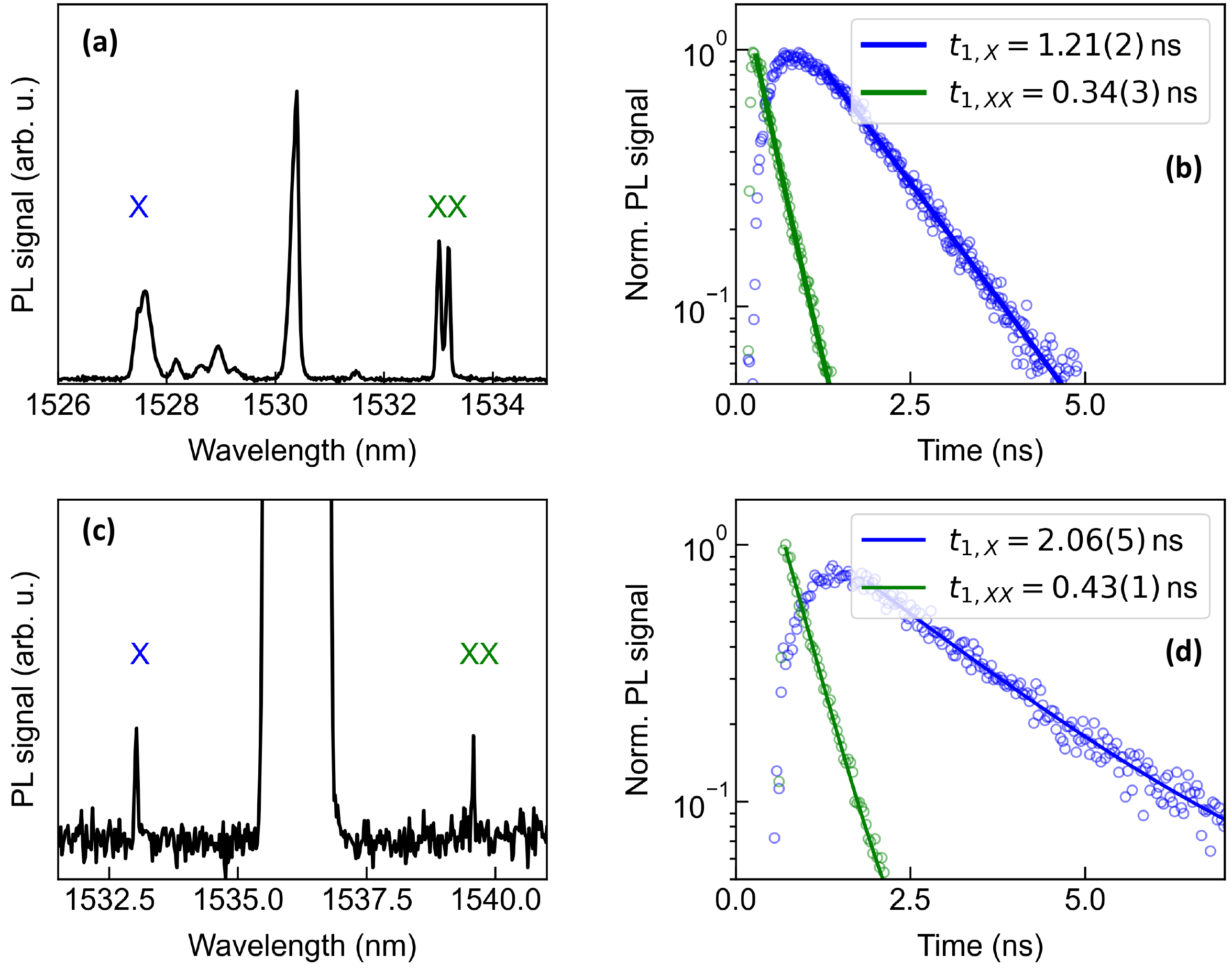}
    \caption{Investigation of a second QD under TPE: (a,b) Emission spectrum and lifetime measurement for the QD from main text for comparison. (c) Emission spectrum and lifetime measurement of another QD from the same sample. (d) Time-resolved measurement of the emission of each QD state under TPE, confirming a 4-times faster decay of the XX state relative to the X-state.}
    \label{fig:SI_other-dot}
\end{figure}

This section presents additional data for another QD from the same sample excited via TPE (see Figure \ref{fig:SI_other-dot}). The emission of this QD is centered around 1535$\,$nm within the telecom C-band. Figure \ref{fig:SI_other-dot}(c) shows the emission spectrum under TPE and the assignment to the respective QD states. The spectral fingerprint of the QD is similar to the one presented in the main article (cf. Figure \ref{fig:SI_other-dot}(a)), but smaller emission linewidths below 30$\, \upmu \text{eV}$ are observed. Note that no additional notch filter was used to suppress the reflected laser in this measurement, resulting in intense laser scattering. The lifetime measurements confirm a 4-times faster decay of the XX-state relative to the X-state (see Figure \ref{fig:SI_other-dot}(d)), rendering this type of QD interesting for simultaneously achieving high photon-indistinguishabilities and entanglement fidelities in future work.

\section{Evaluation of Preparation Fidelity}
\subsection{Rabi Fits vs Cross-Correlation}
\label{sec:PrepFidelity}
The observation of Rabi rotations in the integrated emission intensity as a function of the excitation power, i.e. pulse area, observed in Figure 2(a,b) of the main article is a signature for the coherent population inversion when using resonant excitation schemes. To extract the preparation fidelity $\mathcal{F_{\text{prep}}}$ for the upper state quantitatively, we use two different and independent approaches in our work: (1) By fitting the integrated emission intensity as a function of the excitation power using a theoretical model for the Rabi rotations and (2) by performing photon cross-correlation measurements between XX and X photons as a function of the excitation power. As summarized in SI-Table \ref{tab:compare} the preparation fidelities extracted using these two approaches agree within the standard error.
\begin{table}[ht]
\caption{Comparison of extracted TPE preparation fidelities using different methods}
\label{tab:compare}
\begin{tabular}{l|l|l}
Method  & Preparation Fidelity & Fit Residual \\ \hline
Analytical approx. & $82(4)\%$            & 0.055        \\
XX-X Cross-Corr.   & $81(2)\%$            & -                 
\end{tabular}
\label{tab}
\end{table}

For extracting the preparation fidelity from the integrated emission intensity vs. excitation power, we followed the adapted analytical model used in the Supplemental Material of Ref. \onlinecite{pennacchietti2023oscillating}. The model uses an approximated analytical expression for the Rabi oscillations of a 2-level-system in the time-domain under continuous excitation\cite{fox2006quantum} extended to be applicable for pulsed excitation. The obtained expression describes the occupation probability of the excited state $|c_e|^2$ during the Rabi rotations as a function of pulse area $\Theta$, which for a fixed pulse duration translates into a excitation-power dependence. With the fitting parameter $\xi$, i.e., the damping rate normalized by the Rabi frequency, the model reads 
\begin{equation}
    \left|c_e\right|^2=\frac{1}{2\left(1+2 \xi^2\right)}\left(1-\left(\cos (\Theta)+\frac{3 \xi}{\sqrt{4-\xi^2}} \sin (\Theta)\right) e^{-3 \Theta \xi / 2}\right) ,
\end{equation}
which is used for the fits presented in Figure 2(a,b) of the main article. A preparation fidelity for the XX state of $82(4)\%$ is obtained by extrapolating the exponential envelope function.

To independently verify the result from the Rabi fit, we extracted the preparation fidelity in a second approach via photon cross-correlation measurements between XX and X photons, by comparing the integrated coincidences originating from the same cascade to those of photons from different cascades \cite{wang2019demand,rota2020entanglement,neuwirth2022multipair}. Such a cross-correlation histogram is depicted in Figure \ref{fig:SI_fidelity-extraction}(a)) for an excitation power corresponding to a $\pi$-pulse. In this measurement, an increasing ratio of center to side peak area corresponds to a decreasing preparation fidelity, i.e., not every laser pulse results in a complete population inversion, leading to bunching in the coincidence histogram. Integrating over all coincidences within the full laser repetition period of $12.5\,$ns (cf. Figure \ref{fig:SI_fidelity-extraction}(c)), the preparation fidelity is extracted as:
\begin{equation}
    \mathcal{F}_{\text{prep}}=\frac{A_{\text{side}}}{A_{\text{center}}} \cdot C_{\text{Pol}}
    \label{eq.prep.cross}
\end{equation}
with $A_{\text{side}}$ the average of the integrated coincidences in each of the side peaks, $A_{\text{center}}$ the total number of coincidences in the center peak, and $C_{\text{Pol}}$ a correction factor accounting for polarized detection. Concerning the latter, there are some subtleties to consider:
\begin{itemize}
    \item As we are using a cross-polarized excitation and detection scheme to suppress reflected laser light, the bunching magnitude is $2\times$ larger than without polarization filtering if the horizontally and vertically polarized decay-channels are equally probable (unpolarized source), which needs to be corrected by the re-scaling factor $C_{\text{Pol}}$ as done in Ref. \onlinecite{wang2019demand}. In our case, however, the emission of the QD is partially polarized ($\textrm{DOP}=33\%$, cf. Figure \ref{fig:SI_identification}(a)), counteracting the effect of polarization filtering in our setup. As a first-order approximation, one may account for a finite DOP by choosing $C_{\text{Pol}} = 2-\text{DOP}$,
    resulting in a fully compensated correction factor $C_{\text{Pol}}=1$ for a fully polarized cascade ($\text{DOP}=1$). As shown further below (see Figure \ref{fig:SI_simulate-fidelity-extraction} and associated discussion), simulations reveal a more accurate correction factor of $ C_{\text{Pol}}=1.53$ for our case, which we used to correct the center area (cf. Figure \ref{fig:SI_fidelity-extraction}(d)).
    \item Moreover, the QD investigated in this work shows a blinking effect, additionally masking the analysis of the preparation fidelity according to Equation \ref{eq.prep.cross}. For this reason, we additionally corrected for the blinking by fitting the peak integrals with a double-sided exponential decay (excluding the center coincidence peak), to re-scale the histogram (see Figure \ref{fig:SI_fidelity-extraction}(e)). This blinking correction was applied for each cross-correlation histogram individually before extracting the preparation fidelity via Eq. \ref{eq.prep.cross}. Interestingly, evaluating the blinking magnitude as a function of the pulse area, we find that the relative amount of blinking oscillates in phase with the Rabi rotations, while the timescale of the blinking remains constant (cf. Figure \ref{fig:SI_fidelity-extraction}(b), insets). This can be understood intuitively, as for example a $2\pi$-pulse results in excitation and de-excitation within one laser pulse, thus any blinking cancels out, while at $\pi$-power all blinking can be observed.
    \begin{figure}[ht]
    \centering
    \includegraphics[width=\textwidth]{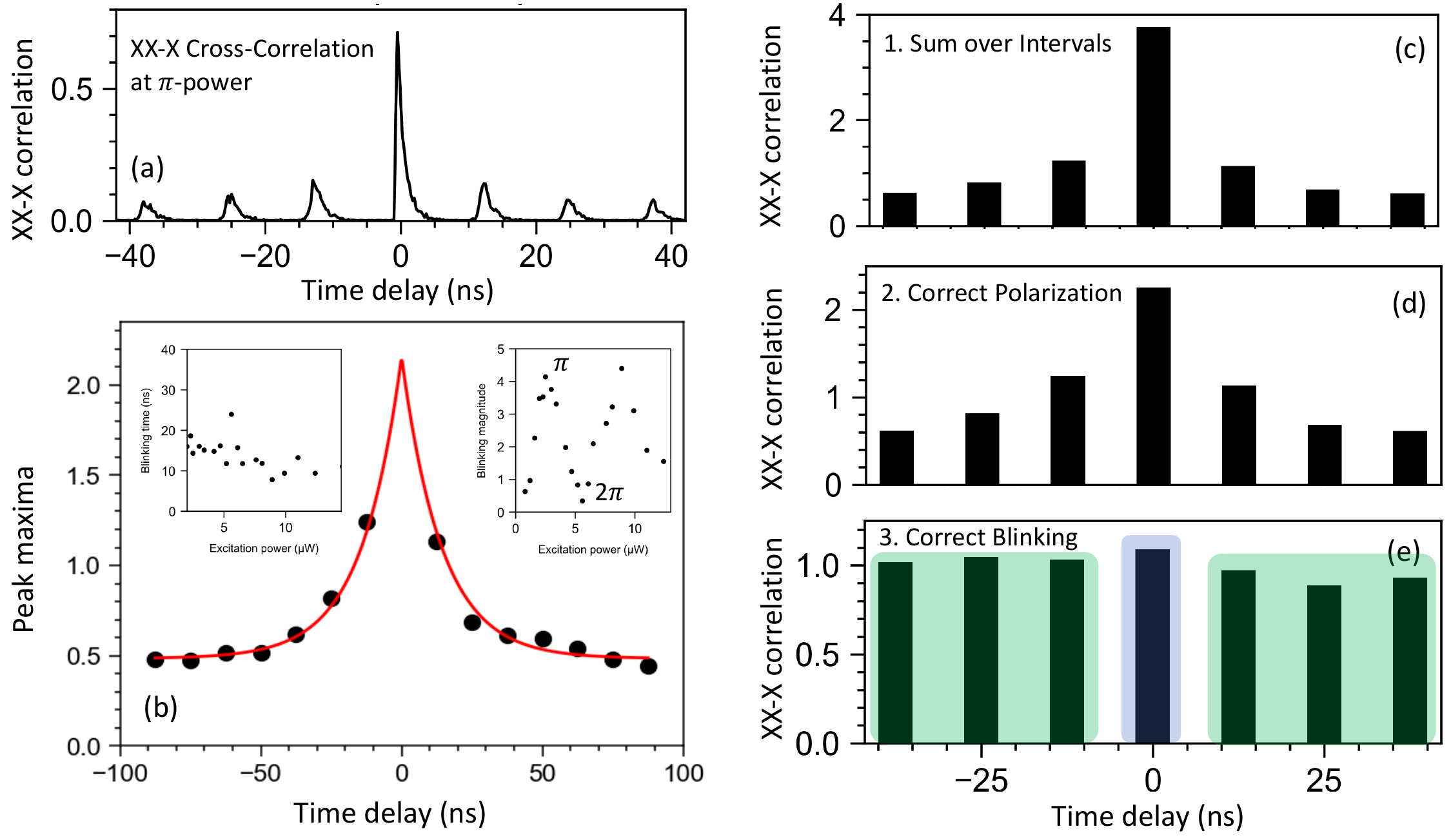}
    \caption{Data evaluation workflow for extracting the preparation fidelity from photon cross-correlation measurements: (a) As-measured XX-X photon cross-correlation histogram under TPE at the $\pi$-pulse ($2.5\,\mu$W). (b) Analyzing the blinking effect by fitting a two-sided exponential to the side-peak maxima. Insets: Extracted blinking timescale (left) and blinking magnitude $m$ (right) as a function of the excitation power. To extract the preparation fidelity the following three steps are applied: (c) the coincidences of each peak are integrated within a full repetition period, (d) the center peak is corrected for polarization effects, and (e) the histogram is re-scaled to correct for the blinking. Finally, the preparation fidelity is extracted as the ratio of the average side peak area (green shaded) and the center peak area (blue shaded).}
    \label{fig:SI_fidelity-extraction}
    \end{figure}
    \end{itemize}
    
\begin{figure}[ht]
    \centering
    \includegraphics[width=\textwidth]{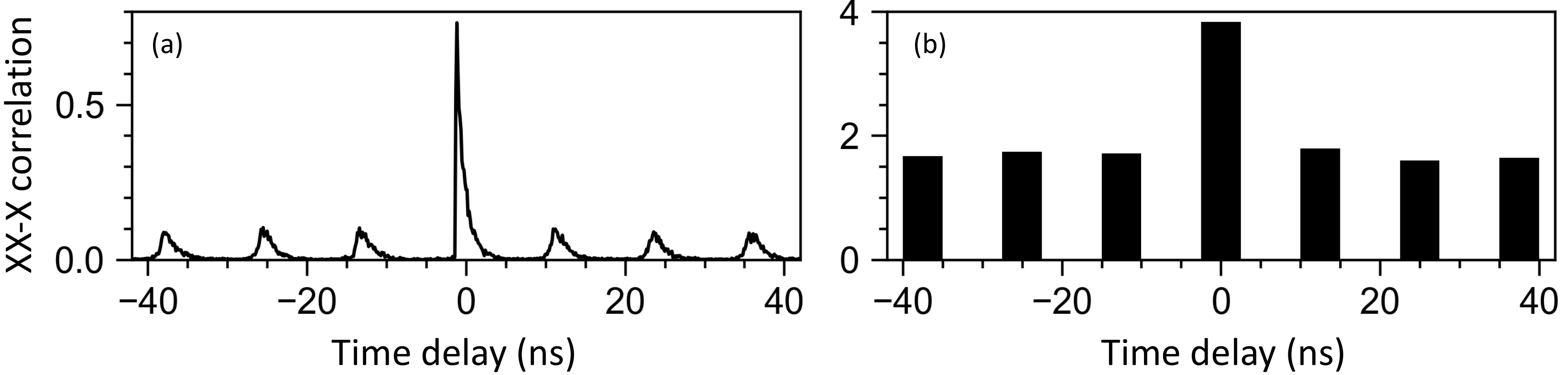}
    \caption{Extracting the preparation fidelity under pulsed above-barrier excitation: (a) Raw XX-X photon cross-correlation histogram and (b) integrated coincidences per peak after polarization- and blinking-correction for extraction of the preparation fidelity.}
    \label{fig:SI_above-barrier-fidelity-extraction}
\end{figure}

Note: Correcting for the blinking effect above, even a preparation fidelity $\mathcal{F}_{\text{prep}}=1$ does strictly not correspond to an on-demand single photon source, as the QD emission is still subject to a random switching between the on- and off-state. We are however confident that such blinking can be controlled by advanced material and device engineering using, e.g., interface passivation or electrical gating.
\\
Applying the analysis discussed above for the experimentally obtained XX-X photon cross-correlation measurements yields a maximum preparation fidelity of $81(2)\%$ under TPE (cf. SI-Table \ref{tab} and Figure 3(d) in main article). 
\\
To gain further insights in how the excitation scheme affects the preparation fidelity, we additionally performed XX-X photon cross-correlation experiments using pulsed above-barrier excitation (at 1470$\,$nm) at saturation power of the X state ($5\,\mu$W). From the results displayed in Figure \ref{fig:SI_above-barrier-fidelity-extraction}(a) and (b), we obtain a preparation fidelity of $44(2)\%$, clearly confirming the sensitive and positive impact of coherent excitation.

\subsection{Additional Simulations of Photon Cross-Correlations}
\label{sec:simulation-cross-correlation}
\begin{figure}[ht]
    \centering
    \includegraphics[width=\textwidth]{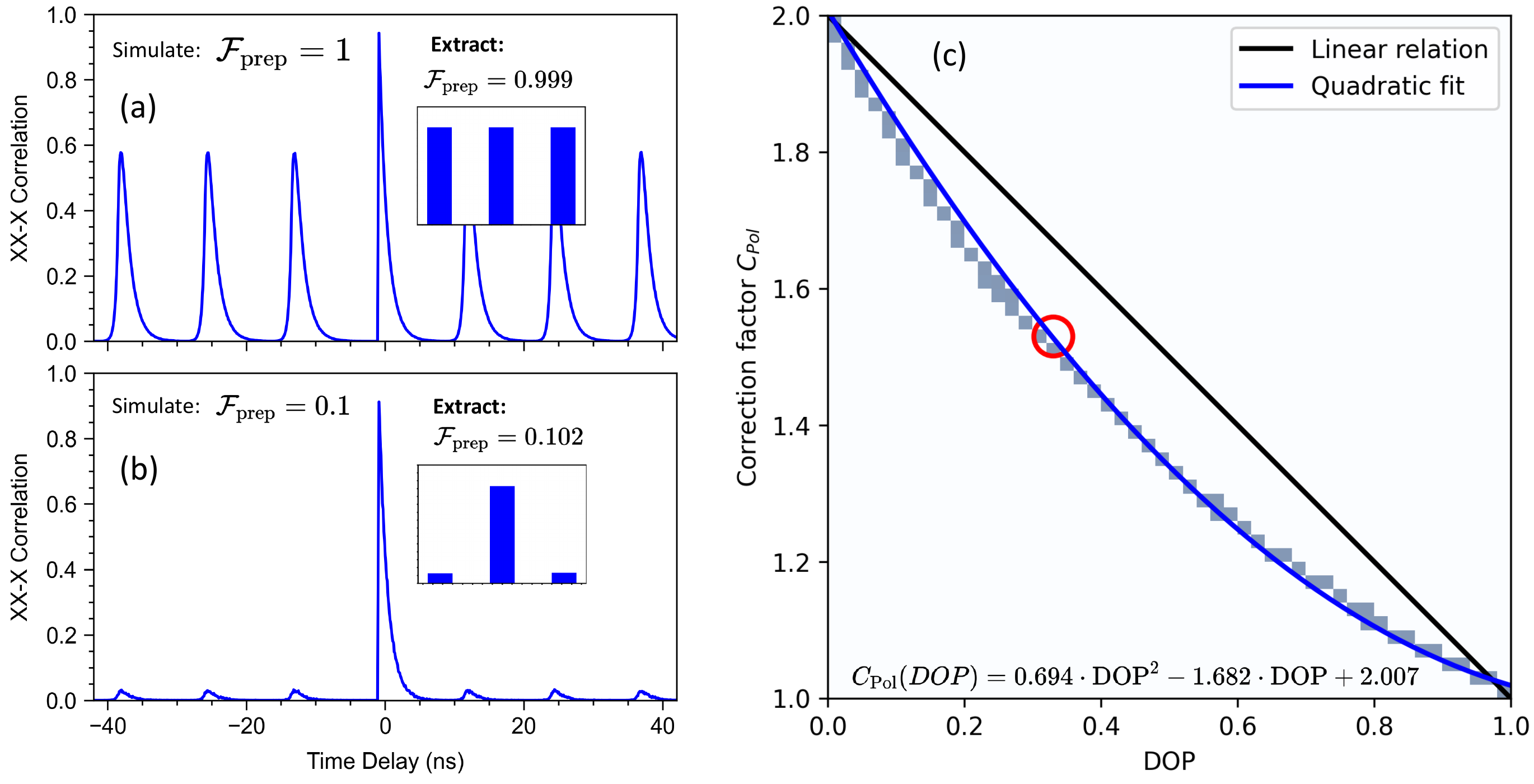}
    \caption{Monte Carlo simulations of the XX-X photon cross-correlation: (a) and (b) Exemplary histograms simulated for perfect ($\mathcal{F}_{\text{prep}}=1$) and low ($\mathcal{F}_{\text{prep}}=0.01$) preparation fidelity (excluding polarization effects). Insets in (a) and (b): Applying the data evaluation workflow from Figure \ref{fig:SI_fidelity-extraction}(c-e) we retrieve the $\mathcal{F}_{\text{prep}}$-values in very good approximation. (c) Correction factor $C_{\text{Pol}}$ extracted from the simulations by accounting for the polarization effects discussed in the text. For our experimental conditions of $\textrm{DOP}=33\%$ a correction factor of $C_{\text{Pol}}=1.53$ is deduced (red circle) and applied to extract the preparation fidelity from experimental results in Figure 3(d) of the main article.}
    \label{fig:SI_simulate-fidelity-extraction}
\end{figure}

This section provides simulation results enabling the non-trivial polarization correction discussed in SI, Section \ref{sec:PrepFidelity}. To find the relation between the measured DOP of the QD under study and the required correction factor $C_{\text{Pol}}$ in Equation \ref{eq.prep.cross}, we performed Monte Carlo simulations of the XX-X radiative cascade accounting for the experimentally determined decay times. The simulations directly incorporate the preparation fidelity, i.e., the probability that a laser pulse excite the XX state. The simulation randomly draws 
 between the two possible polarization decay channels (HH or VV) of the cascade, but keeps only results from one decay channel (e.g. horizontal polarizations) to account for the polarization filtering in our experimental setup. Additionally the simulations account for the known experimental imperfections (detector timing jitter, dead time, setup efficiency, etc.), to generate data sets for a representative comparison with our experimental data. Figure \ref{fig:SI_simulate-fidelity-extraction}(a) and (b) shows the simulated cross-correlation histograms for ideal ($\mathcal{F_{\text{prep}}}=1$) and low ($\mathcal{F_{\text{prep}}}=0.1$) preparation fidelity for the case of unpolarized emission and without polarization-filtering, respectively. The simulated histograms adequately reproduce the details (coincidence peak asymmetries and area ratios) expected for the QD XX-X cascade. Applying the evaluation workflow presented in Figure \ref{fig:SI_fidelity-extraction}(c-e) as a cross-check, we retrieve the preparation fidelities originally fed in the simulations in very good approximation. Finally, including the polarization filtering in our simulation and varying the initial DOP, we are able to extract the polarization correction factor $C_{\text{Pol}}$ reproducing the correct preparation fidelity for each case (cf. Figure \ref{fig:SI_simulate-fidelity-extraction}(c)). The resulting non-linear relation, can be approximated by a quadratic fit, from which we retrieve a polarization correction factor of 1.53 for our case of $\textrm{DOP}=33\%$ (red circle).

\section{Evaluation of Time-Resolved Measurements}
\subsection{Extracting Purity from Second-Order Auto-Correlation Measurement}
\label{sec:HBT-fit}
\begin{figure}[ht]
    \centering
\includegraphics[width=\textwidth]{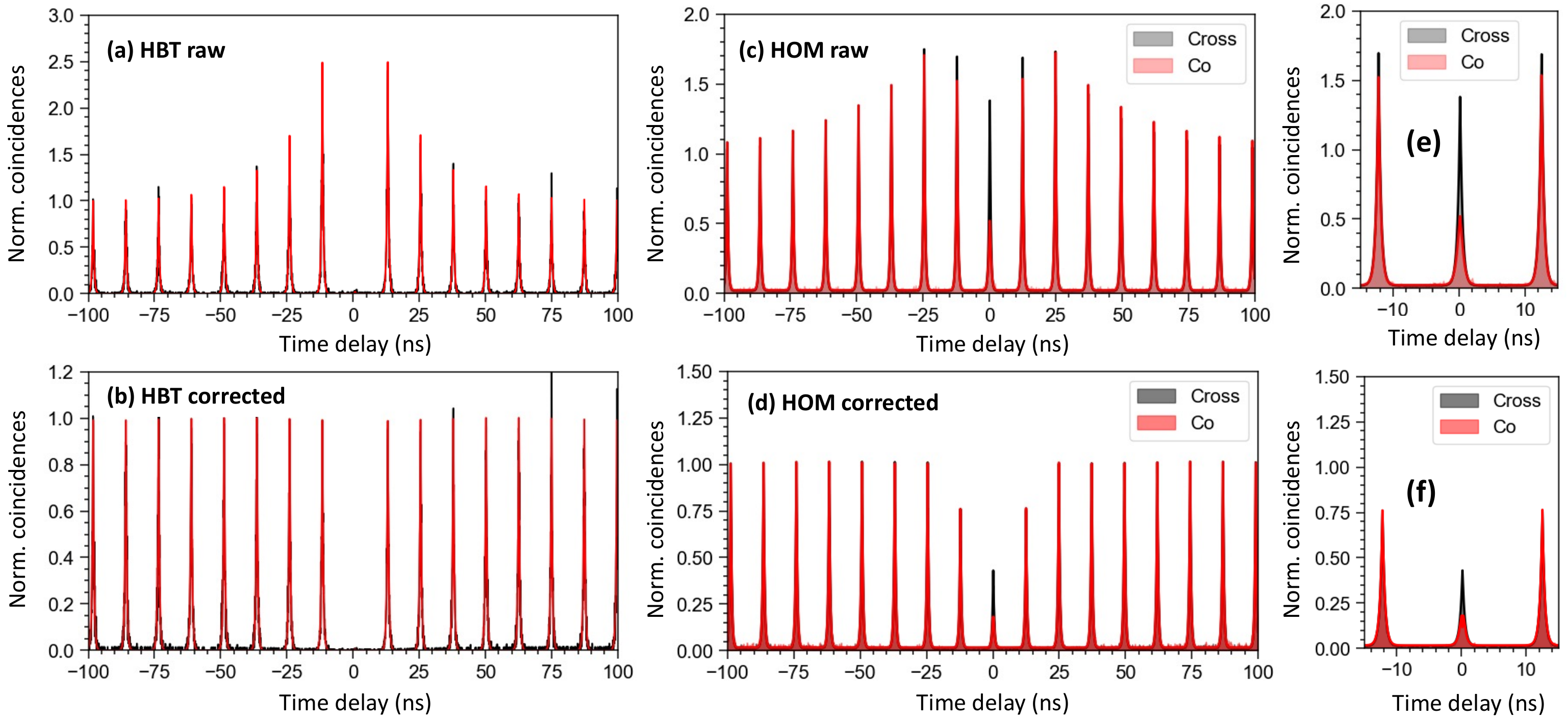}
    \caption{Blinking correction of correlation measurements: (a,b) $g^{(2)}$-data without (top) and with (bottom) blinking correction. Data (black) is shown together with fit according to Eq. \ref{eq:g2} (c,d) Correcting blinking for the HOM data reproduces the expected peak heights for HOM experiments with 12.5$\,$ns delay. Cross-polarized (grey) and co-polarized (red) data is shown together with the fit according to Eq. \ref{eq:HOM}. (e,f) close-up of HOM data without and with blinking correction.}
\label{fig:SI_blinking}
\end{figure}

As explained in the main text, the auto-correlation data was fitted with a sum of two-sided mono-exponential decays. We do not apply any background subtraction nor de-convolution with the system's response function, as the measured decay times are about one order of magnitude slower then the average system response time. We do however account for the blinking effect, by incorporating an exponential envelope function in our fit model as follows
\begin{equation}
C(\tau) = \left[g^{(2)}(0) \cdot e^{-|\tau| / \tau_1} +\sum_{n \neq 0} e^{ -\left|\tau-n \tau_0\right| / \tau_1} \right] \times \,  C_0 \cdot \left( 1+m \right) \cdot e^{-|\tau|/\tau_{\mathrm{blink}}} \hspace{5mm}.
\label{eq:g2}
\end{equation}
Here, the normalized coincidences $C(\tau)$ as a function of the detection time delay $\tau$ are modelled using the zero-delay coincidences $g^{(2)}(0)$, the radiative decay time $\tau_1$, the laser pulse spacing $\tau_0$, a normalization factor $C_0$, a blinking timescale $\tau_{\text{blink}}$ and the blinking magnitude $m$.
The respective best values and standard errors for the parameters are extracted from fits to the experimental data for the X- and XX-state shown in Figure 2(d,e) of the main article (see SI-Table \ref{tab:g2-fit}). Note, that while for some QDs on this sample, adding some above-barrier light reduced the blinking (indicative of a stabilization of the charge environment), this has not been the case for the QDs studied here. This is in line with the observation that additional above-barrier support increases the amount of excess charges instead of saturating them (cf. SI section \ref{sec:TPE-plus-above-barrier}).
\\
\begin{table}[ht]
    \centering
    \caption{Fitting parameters for $g^{(2)}$-measurement of XX and X photons under TPE}
    \begin{tabular}{c|c|c} 
        Parameter & X Fit & XX Fit \\
         \hline
        decay time  $\tau_1$ & 1.44(1)$\,$ns & 0.36(1)$\,$ns \\
        blinking time $\tau_{\text{blink}}$ & 16.9(3)$\,$ns & 16.7(3)$\,$ns \\
        blinking magnitude $m$ & 4.96(7) & 3.19(4) \\
        On/Off-times $\tau_{\text{on}}$ /  $\tau_{\text{off}}$ & 20$\,$ns / 100$\,$ns & 22$\,$ns / 70$\,$ns \\
        Quantum Efficiency $QE$ & 17$\,\%$ & 23$\,\%$  \\
        center-peak contribution $g^{(2)}(0)$ & 0.015(4) & 0.005(4) \\
    \end{tabular}
    \label{tab:g2-fit}
\end{table}
The blinking can be used to obtain information about the on/off-ratio as well as a lower bound for the quantum efficiency $QE$. It is a lower bound as random jumping of the emission energy relative to the detected energy can cause on/off-switching that leads to blinking in the auto-correlation measurements reducing the extracted $QE$, even though the QD still emits a photon. According to Ref.$\,$\onlinecite{santori2001triggered}, the blinking can be heuristically described by the height $h$ of the $m$-th. peak maximum obeying
$ h_{m \neq 0} = 1+\frac{\tau_{\mathrm{off}}}{\tau_{\mathrm{on}}} \cdot e^{-\left(\frac{1} { \tau_{\mathrm{off}}}+\frac{1}{\tau_{\mathrm{on}}} \right) \cdot |m \tau_0|}$
where $\tau_{\text{on}}$ and $\tau_{\text{off}}$ are the on-/off-times of the emitter. This relates to the fitting parameters of our auto-correlation fit (Eq. \ref{eq:g2}) via
$\tau_{\text{on}} = \left(\frac{1}{1+m} \right)\tau_{\text{blink}} \, , \,\,\, \tau_{\text{off}} = \left(1+m\right) \tau_{\text{blink}}$
and thus the quantum efficiency can be estimated via
$QE = \frac{\tau_{\text{on}}}{\tau_{\text{on}}+\tau_{\text{off}}} = \frac{1}{1+m} .$
Using this model, results in an extracted quantum efficiency around $20\%$.
\\
As mentioned in the main text, the extracted blinking envelope function can be used to correct the data for blinking by dividing the data by $\left( 1+m \right) \cdot e^{-|\tau|/\tau_{\mathrm{blink}}}$. This can be done for the auto-correlation of second order, as well as for the HOM data, which yields blinking-corrected histograms as shown in Figure \ref{fig:SI_blinking}. This is especially useful to confirm that the expected relative ratios of the peak areas in the HOM experiment are reproduced.

\subsection{Extracting Indistinguishability from HOM Measurement}
\label{sec:HOM-fit}
To extract the two-photon interference visibility as a measure for the photon indistinguishability from Hong-Ou-Mandel (HOM) measurements, different methods can be applied. Using only the co-polarized measurement data and comparing center to side peak areas is not precise in the presence of blinking. Hence, to evaluate our HOM measurements we compare the co-polarized measurement with a maximally distinguishable cross-polarized reference measurement. In order to compare both measurements, they are fitted according to equation
\begin{eqnarray*}
C_{\text{co}}(\tau) =&& \left[ A  \cdot e^{-|\tau| / \tau_1} \left( 1-V_{\text{PS}} e^{-|\tau|/\tau_2} \right) + \sum_{n \neq -1,0,1} e^{ -\left|\tau + n \tau_0\right| / \tau_1}  
+  \frac{3}{4} \left(e^{ -|\tau + \tau_0 | / \tau_1} +  e^{ -|\tau- \tau_0| / \tau_1} \right) \right] \\
&&\times  C_0 \cdot \left( 1+m \right) \cdot e^{-|\tau|/\tau_{\mathrm{blink}}}
\label{eq:HOM}
\end{eqnarray*}
for the co-polarized data and 
\begin{eqnarray*}
C_{\text{cross}}(\tau) =&& \left[ A \cdot e^{-|\tau| / \tau_1} + \sum_{n \neq -1,0,1} e^{ -\left|\tau + n \tau_0\right| / \tau_1} + \frac{3}{4} \left(e^{ -|\tau + \tau_0 | / \tau_1} +  e^{ -|\tau- \tau_0| / \tau_1} \right) \right] \\
&& \times  C_0 \cdot \left( 1+m \right) \cdot e^{-|\tau|/\tau_{\mathrm{blink}}}
\end{eqnarray*}
for the corresponding cross-polarized data.
Here $A$ captures the area of the center peak, $V_{\text{PS}}$ is the post-selected visibility value that is set to zero for fitting the cross-polarized data, $\tau_1$ the decay time and $\tau_2$ the coherence time. The blinking is included in the same way as in the $g^{(2)}$-fit (Eq. \ref{eq:g2}) with the blinking amount $m$ and timescale $\tau_{\text{blink}}$ and a normalization factor $C_0$. Note that we do not subtract any background or correct for limited purity here as the $g^{(2)}$-measurement confirmed the good suppression of the reflected laser light. After fitting both co- and cross-polarized data, the data is divided by the normalization constant $C_0$, which ideally should be 1, but insufficient statistics sometimes leads to imperfect normalization. In this way both measurements are normalized to their respective Poisson levels. The fitting parameters of the HOM fit of the XX photons for co- and cross-polarized measurement are collected in SI-Table \ref{tab:hom-fit}. \\

\begin{table}[ht]
    \centering
     \caption{Fitting parameters for HOM-measurement of XX photons}
    \begin{tabular}{c|c|c} 
        Parameter & Cross-polarized & Co-polarized \\
         \hline
        normalization constant $C_0$ & 1.000(1) &  0.999(1) \\
        decay time  $\tau_1$ & 0.36(1)$\,$ns & 0.36(1)$\,$ns \\
        blinking time $\tau_{\text{blink}}$ & 23(1)$\,$ns & 32(2)$\,$ns  \\
        blinking contribution $m$ & 2.05(5) & 1.41(5) \\
        post-selected indistinguishability $V_{\text{PS}}$ & - & 73(6)$\%$ \\
        coherence time $\tau_2$ & - & 0.15(5)$\,$ns \\
        Normalized integrated coincidences $A_{\text{4ns}}$ & 3.88 & 2.48
    \end{tabular}
    \label{tab:hom-fit}
\end{table}

\begin{figure}[ht]
    \centering
    \includegraphics[width=\textwidth]{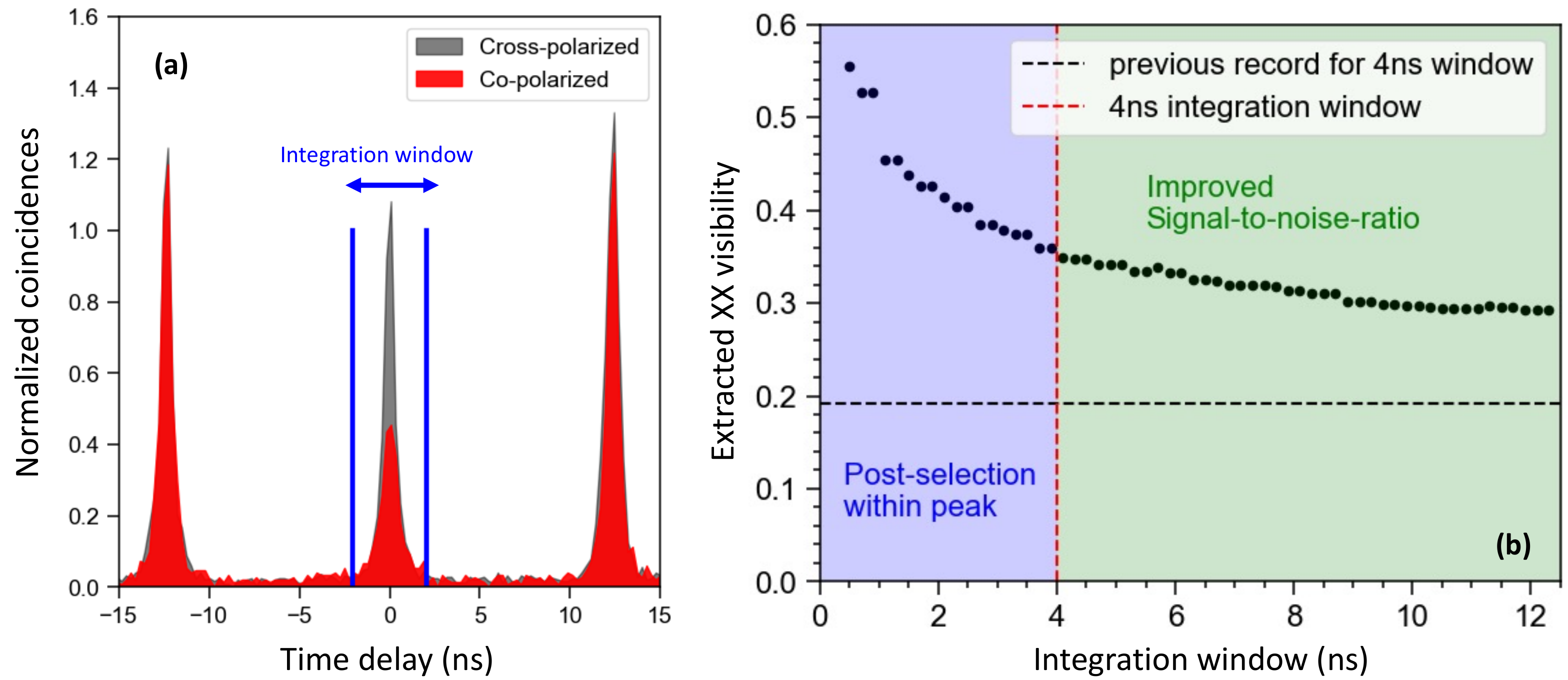}
    \caption{(a) The majority of the coincidence data lies within a 4$\,$ns time window. (b) Different integration time windows lead to different extracted $V$-values as for windows $> 4\,$ns the background emission contributes and for windows $< 4\,$ns the data is post-selected. For all integration windows we improve over the previous state-of-the-art for C-band QDs (black dashed line) \cite{holewa2023scalable}.}
    \label{fig:SI_hom}
\end{figure}

After confirming the normalization of the date by finding $C_0=1$, the coincidences within a 4$\,$ns time window for the co- as well as for the cross-polarized data are integrated and the visibility is computed as $V_{\text{HOM}} = 1- \frac{A_{\text{co}}}{A_{\text{cross}}}$. The size of the integration window influences the value of $V_{\text{HOM}}$. As discussed in the main text, using a reduced integration window instead of the full laser period of 12.5$\,$ns improves the signal-to-noise-ratio, especially as the count rate was relatively low in our experiments, but does not correspond to any post-selection yet as about 99$\%$ of the coincidence data lie within the 4$\,$ns integration window (see Figure \ref{fig:SI_hom}(a)). Thus, only reducing the window further than 4$\,$ns improves $V_{\text{HOM}}$ super-linearly via post-selection. Therefore, we state also the 4$\,$ns integration window in the main article, to also compare better to previous works using the same integration window. \\
The effect of changing the integration time window is illustrated in Figure \ref{fig:SI_hom}(b), where the transition from the regime in which only the signal-to-noise-ratio is improved to the post-selection regime can be readily observed. We have indicated the 4$\,$ns time window (black dashed line) and the previous record of InP indistinguishability obtained with that integration time window\cite{holewa2023scalable} (red dashed line), that is clearly surpassed by using coherent excitation.

\clearpage
\putbib
\end{bibunit}

\end{document}